\documentclass[twoside]{article}
\usepackage{amsfonts}
\usepackage{stmaryrd}
\usepackage{amssymb}
\usepackage{euscript}
\usepackage{amsthm}
\usepackage{amsmath}
\usepackage{amscd}
\usepackage{latexsym}
\usepackage{mathrsfs}
\usepackage{graphicx}
\usepackage{color}
\usepackage{dsfont}
\usepackage{bm}
\usepackage{hyperref}
\usepackage{soul}\setstcolor{red}

\numberwithin{equation}{section}

\newtheorem{theorem}{Theorem}[section]
\newtheorem{lem}{Lemma}[section]
\newtheorem{pro}{Proposition}[section]
\newtheorem{cor}{Corollary}[section]
\newtheorem{rem}{Remark}[section]
\newtheorem{rems}{Remarks}[section]
\newtheorem{ex}{Example}[section]
\newtheorem{defi}{Definition}[section]
\newtheorem{hyp}{Assumption}[section]
\newtheorem{con}{Conjecture}[section]

\newcommand{\sect}{\section}
\newcommand{\ssc}{\subsection}
\newcommand{\sssc}{\subsubsection}

\newcommand{\bt}{\begin{theorem}}
\newcommand{\et}{\end{theorem}}
\newcommand{\bl}{\begin{lem}}
\newcommand{\el}{\end{lem}}
\newcommand{\bp}{\begin{pro}}
\newcommand{\ep}{\end{pro}}
\newcommand{\bcor}{\begin{cor}}
\newcommand{\ecor}{\end{cor}}
\newcommand{\bcon }{\begin{con} \rm }
\newcommand{\econ }{\end{con}}
\newcommand{\lab }{\label }
\newcommand{\bd}{\begin{defi} \rm }
\newcommand{\ed}{\end{defi}}
\newcommand{\brem }{\begin{rem} \rm }
\newcommand{\erem }{\end{rem}}
\newcommand{\brems }{\begin{rems} \rm }
\newcommand{\erems }{\end{rems}}
\newcommand{\bhyp }{\begin{hyp} \rm }
\newcommand{\ehyp }{\end{hyp}}
\newcommand{\bex}{\begin{ex} \rm }
\newcommand{\eex}{\end{ex}}
\newcommand{\be}{\begin{equation}}
\newcommand{\ee}{\end{equation}}
\newcommand{\bde}{\begin{displaymath}}
\newcommand{\ede}{\end{displaymath}}
\newcommand{\beq}{\begin{eqnarray*}}
\newcommand{\eeq}{\end{eqnarray*}}
\newcommand{\beqa}{\begin{eqnarray}}
\newcommand{\eeqa}{\end{eqnarray}}
\newcommand{\bea}{\begin{align*}}
\newcommand{\eea}{\end{align*}}
\newcommand{\Exx}[2]{\mathbb{E}^r_{#1}\!\left[\,#2\,\right]}

\newcommand{\ind}[1]{\I_{\{#1\}}}

\def\proof{\noindent {\it Proof. $\, $}}
\def\endproof{\hfill $\Box$ \vskip 5 pt}

\newcommand{\lgd}{\mbox{L}}
\newcommand{\lgdE}{\mbox{L}_E}
\newcommand{\tauE}{\tau_E}

\def\I{\mathds{1}}
\def\wh{\widehat}
\def\wt{\widetilde}
\def\phi{\varphi }

\newcommand{\epsiQ}{Q}
\newcommand{\Vtop}{\pi}
\newcommand{\Pitoa}{\Pi }

\newcommand{\pC}{C}

\newcommand{\pCc}{c}

\newcommand{\ACtheta}{A^{C,R}}
\newcommand{\recR}{R}
\newcommand{\wtAC}{\widetilde{A}^{C}}
\newcommand{\wtA}{\widetilde{A}}
\newcommand{\wtC}{\widetilde{C}}

\newcommand{\Blr}{B^{l}}
\newcommand{\Bbr}{B^{b}}
\newcommand{\Bf}{B^f}

\newcommand{\rlb}{f}
\newcommand{\rll}{f^{l}}
\newcommand{\rbb}{f^{b}}
\newcommand{\psif}{\psi^f}
\newcommand{\pdj }{\kappa^j}

\newcommand{\bartau }{\widehat{\tau }}

\newcommand{\rcl}{c^{l}}
\newcommand{\rcb}{c^{b}}

\newcommand{\Bilr}{B^{i,l}}
\newcommand{\Bibr}{B^{i,b}}

\newcommand{\rilb}{h^{i}}
\newcommand{\rill}{h^{i,l}}
\newcommand{\ribb}{h^{i,b}}

\def\t1{\tau_{(1)}}

\def\rr{\mathbb R}
\def\ff{{\mathbb F}}

\def\gg{{\mathbb G}}

\def\F{{\cal F}}

\def\G{{\cal G}}

\def\P{\mathbb P}

\def\Q{\mathbb Q}

\def\E{{\mathbb E}}

\def\EQh{{\mathbb E}_{{\mathbb Q}^h}}

\def\FVA{\textrm{FVA}}
\def\FCA{\textrm{FCA}}
\def\FBA{\textrm{FBA}}

\def\CVA{\textrm{CVA}}
\def\DVA{\textrm{DVA}}
\def\DVAF{\textrm{DVA}^{f}}
\def\DVAFn{\textrm{DVA}^{{f},-}}
\def\DVAFp{\textrm{DVA}^{{f},+}}
\def\CVAF{\textrm{CVA}^{f}}
\def\LVA{\textrm{LVA}}

\newcommand{\Keywords}[1]{\par\noindent{\small{\bf Keywords\/}: #1}}
\newcommand{\Class}[1]{\par\noindent{\small{\bf Mathematics Subjects Classification (2010)\/}: #1}}

\title{{\Large \bf RISK-NEUTRAL VALUATION UNDER DIFFERENTIAL FUNDING COSTS, DEFAULTS AND COLLATERALIZATION}\vskip 65 pt}

\author{Damiano Brigo\footnote{The research of Damiano Brigo and Marek Rutkowski was supported by the EPSRC Mathematics Platform Grant EP/I019111/1 {\it Mathematical Analysis of Funding Costs} at Imperial College London.}\\Dept. of Mathematics \\ Imperial College London \and Cristin Buescu \\ Dept. of Mathematics \\ King's College London \\ \and Marco Francischello \\ Dept. of Mathematics \\ Imperial College London \and Andrea Pallavicini \\ Dept. of Mathematics \\ Imperial College London \\ \and Marek Rutkowski\footnote{The research of Cristin Buescu and Marek Rutkowski was supported by the DVC Research Bridging Support Grant {\it BSDEs Approach to Models with Funding Costs.}} \\ School of Mathematics and Statistics \\ University of Sydney} 

\date{\vskip 35 pt \today}

\begin{document}
\maketitle
\vskip 35 pt
\begin{abstract}
	
We develop a unified valuation theory that incorporates credit risk
(defaults), collateralization and funding costs, by expanding the
replication approach to a generality that has not yet been studied
previously and reaching valuation when replication is not assumed.
This unifying theoretical framework clarifies the relationship between
the two valuation approaches: the adjusted cash flows approach
pioneered for example by Brigo, Pallavicini and co-authors (\cite{BFP16,BFP17,PPB12}) and the classic replication approach illustrated for example by
Bielecki and Rutkowski and co-authors (\cite{BCR2017,BR15}). In particular, results
of this work cover most previous papers where the authors studied
specific replication models.


\vskip 20 pt
\Keywords{risk-neutral valuation, replication, funding costs, default, collateral}
\vskip 20 pt
\Class{91G40,$\,$60J28}
\end{abstract}



\newpage

\sect{Introduction}  \label{sec1}

The recent financial crisis has brought to prominence the adverse impact that credit risk can have not only to the overall volume of market trades, but to the viability and existence of significantly large market players. This required modifying the standard valuation procedure of individual contracts, and this modification was often done under the immediate practical pressures on an ad-hoc basis, while focusing on a particular setting and on a specific contract, without
regard for a general theory. We develop here a unified valuation theory that incorporates credit risk (defaults), collateralization and funding costs, by expanding the replication approach to a generality that has not yet been studied (see \cite{BR15} for the replication approach in the case of collateralization and funding costs, but without defaults). This theory covers all papers that study specific replication models.

Prior to the financial crisis of 2007-2008 institutions tended to ignore the credit risk of highly-rated counterparties in valuing and hedging contingent claims traded over-the-counter (OTC), claims which are in fact bilateral contracts negotiated between
two default-risky entities. Then in just the short span of one month of 2008 (Sep 7 to Oct 8) eight mainstream financial institutions experienced critical credit events in a painful reminder of the default-riskiness of even large names (the eight were:
Fannie Mae, Freddie Mac, Lehman Brothers, Washington Mutual, Landsbanki, Glitnir and Kaupthing, to which one could also add Merrill Lynch). One of the explosive manifestations of this crisis was the sudden divergence between the rate of overnight indexed swaps (OISs) and the LIBOR rate, pointing to the credit and liquidity risk existing in the interbank market. This forced dealers and financial institutions to reassess the valuation of OTC claims. The initial reaction, continuing to a large extent these days, has been to introduce a series of additive valuation adjustments, today often referred to with the collective name of XVAs, that would adjust the value of the deal for the costs and risks ignored in previous valuation practices. The problem is that a number of such costs and risks display nonlinear features, so that the additive split is merely an approximation.

In terms of the existing literature, it is difficult to do justice to the previous work on such valuation adjustments,
which intertwines two strands that have been developed in parallel by academics and practitioners. For a full introduction to valuation adjustments and all related references, we refer to the first chapter of either \cite{BMP12} or \cite{CBB2014}. Here we will summarize only the features that are most relevant to this work. Firstly, the {\it credit valuation adjustment} (CVA) corrects the value of a trade with the expected costs borne by the dealer due to scenarios where the counterparty may default. CVA had been around for some time, see for example \cite{BrigoMas}, and in its most sophisticated version can include credit migration and ratings transition, see for example \cite{BC2013}. Secondly, the {\it debit
valuations adjustment} (DVA), which can be interpreted as the CVA seen from the other side of the trade, corrects the price with the expected benefits to the dealer due to scenarios where the dealer might default before the end of the trade. This latter correction may lead to a controversial profit that can be booked when the credit quality of the dealer deteriorates. For example, Citigroup reported in a press release on the first quarter revenues of 2009 that ``Revenues also included [...] a net 2.5\$ billion positive CVA on derivative positions, excluding monolines, mainly due to the widening of Citi's CDS spreads". Accounting standards by the FASB accept DVA, whereas the Basel Committee does not recognize it in the risk measurement space; see, for example, \cite{BMP12} for a detailed discussion. On top of this, DVA is very difficult to hedge, as this would involve selling protection on oneself. The spread risk is therefore hedged via proxy hedging, trading in names that are thought to be correlated to one's own bank, but this does not help with the jump to default risk (see again  \cite{BMP12} for a discussion).

More recently, the {\it funding} and {\it capital valuation adjustments} (abbreviated as FVA and KVA, respectively) were introduced. FVA is the price adjustment due to the cost of funding the trade. Trading desks back the deal with a client by hedging it with
other dealers in the market, and this may involve maintaining a number of hedging accounts in the underlying assets, cash,  or other correlated assets when proxy-hedging.  The funds needed for these operations are typically raised from the internal treasury of the dealer but they ultimately come from external funders. Interest charges on all the related borrowing and lending activities need to be covered, and this is added to the valuation. Michael Rapoport reported in the Wall Street Journal, on Jan 14, 2014, that funding valuation adjustments cost J.P. Morgan Chase \$1.5 billion in the fourth quarter results. More recently, KVA has started being discussed for the cost of capital one has to set aside in order to be able to trade. We will not address KVA here, since even its very definition is currently subject to intense debate in the industry, but we point the reader to \cite{BFPkva} for an indifference pricing approach based on the risk-adjusted return on capital (RAROC).

All such adjustments may concern both over-the-counter (OTC) derivatives trades and derivatives trades done through central counterparty clearing houses (CCP), and we refer to \cite{CCPPallaBr2014} for a comparison of the two cases where the full mathematical structure of the problem of valuation under possibly asymmetric initial and variation margins, funding costs, liquidation delay and credit gap risk is explored, resulting in BSDEs and semilinear PDEs. Several kinds of valuation adjustments are also discussed in \cite{CrepeyZorana}. As mentioned above, the rigorous theory of valuation in the presence of all such effects can be quite challenging; in general, it does not lead to an additive split in the above adjustments, but rather to nonlinear valuation paradigms that are based on advanced mathematical tools, such as semi-linear PDEs or BSDEs (see, e.g., \cite{NR1,NR2,NR4}).
A significant leap was achieved by the ``adjusted cash flows" approach whose practical applicability led to its implementation in the industry.


The comprehensive ``adjusted cash flows'' approach adopted in works by Brigo, Pallavicini and co-authors (\cite{BFP16,BFP17,PPB12}) hinges on modifying the future cash flows of the contract by adding explicit cash contributions representing the additional risks and costs, and then taking risk-neutral expectation of the deflated adjusted cash flows. The resulting general master pricing equation with all effects was initially introduced in Pallavicini et al. \cite{PPB11,PPB12}, and the existence and uniqueness of solutions to such equations was rigorously studied, for example, in \cite{BFP16,BFP17}. This approach extends cash flows to which risk neutral expectation applies, but such extensions do not necessarily result in additive adjustments to basic risk neutral prices, in that from a pathwise point of view the adjusted cash flows depend on the future values of the trade itself. Furthermore, self-financing replicating strategies that are implicitly assumed to underpin the risk-neutral valuation are not written explicitly and are instead merely assumed to exist. Other existing works (see, e.g.,
\cite{BCS17,BK11,C2015a,C2015b, CBB2014,CS2016}) start from the self-financing condition and present explicit calculations, with a varying degree of rigour, leading to particular pricing equations.

The contribution of this paper is to provide a unifying theoretical framework that clarifies the relationship between the two
approaches: the ``adjusted cash flows'' approach and the replication approach. First, we extend the ``adjusted cash flows'' approach by incorporating external funding costs. Then, this generalized ``adjusted cash flows'' approach is reconciled with the classic, albeit typically nonlinear in the present framework, replication paradigm (\cite{BCR2017,BR15}), thereby validating the ``adjusted cash flows'' approach as a sound way to price contracts in practice. Furthermore, we show that for claims that can be replicated, the explicit expression of the cash flows adjustments that need to be operated in the valuation under some martingale measure chosen for the pricing purposes is no longer the result of astute insights into the contract time-line, but an immediate outcome of the replication approach. We also demonstrate via an ``invariance principle'' that the risk-neutral valuation formula, which involves the risk-free rate often associated with a non-tradeable asset, can be cast in the replication approach in a valuation formula that eliminates all dependence on this illusory risk-free rate. For a practical example in this unified framework we refer to \cite{BBR} where all the additional risk factors are integrated in the valuation without forcing separate adjustments to the price. For a specific contract this leads to the modified Black-Scholes pricing formula with dividends, which in turn allows for efficient sensitivity analysis.

The paper is organized as follows. Section \ref{sec2} presents the ``adjusted cash flows'' approach by Brigo, Pallavicini and co-authors, including the margining, default closeout and funding extensions, and expanding it further to include in Section \ref{sec2.7} detailed analysis of the treasury side of the cash flows and thus looking also at the external funding adjustment.
Such analysis shows clearly the nonlinear features affecting the valuation and discusses further under which conditions the valuation formula can be disentangled in a series of additive adjustments. The invariance result stating that the risk-free rate disappears from final valuation equations, previously discussed in \cite{BFP16,BFP17,PPB12}, is examined in detail here and reintroduced in a fully general setting.

Section \ref{sec3} presents a rigorous derivation of the valuation equations under a replication approach with explicit accounts and self-financing conditions, first in the linear case and then when nonlinear effects show up. The treatment provides formal proofs and generalizes a number of results previously obtained through the risk-neutral extended cash flows approach of Section \ref{sec2}.
We emphasize the nonlinear nature of valuation adjustments and we show the natural link between the replication approach and backward stochastic differential equations (BSDEs) for the price and hedge.

Section \ref{sec4} discusses the case of an incomplete market where we relax the replication assumption that had been adopted more implicitly in Section \ref{sec2} and explicitly in Section \ref{sec3}. For the incomplete case, we can still present a derivation that is reminiscent of the previous sections, but which holds under an alternative set of assumptions based on the idea that the extended market should remain arbitrage-free.

\section{Risk-Neutral ``Adjusted Cash Flows'' Approach} \label{sec2}



To introduce the risk-neutral ``adjusted cash flows" approach, we will use a simple model of a bank consisting of the trading desk, the external counterparty of the trade, the treasury of the bank where the trading desk sits, and the external entity providing or receiving funding to/from the treasury. We are now going to explain all the flows that occur in this model. Let the process $(A_t)_{t \in [0,T]}$ represent all promised cash flow of the contract with maturity $T$. By the valuation of the contract, we mean searching for the price the trader is willing to pay to enter the contract with the counterparty.

The starting point is `clean' price of the contract, which is the price of the non-defaultable uncollateralized contract $A$ funded at the risk-free rate. Additional factors like default, collateral and later on funding costs require adjustments to the cash flows.

We will use the following notation:
\begin{itemize}
\item $\Pitoa(t,s,A)$ are the promised (credit-collateral-funding free) cash flows of the contract $A$ from $t$ to $s$, discounted back at time $t$ with the risk-free rate account $B^r$;
\item $\gamma (t,s,C) $ are the discounted cost-of-collateralization cash flows, representing flows of interest remuneration or cost due to collateral $C$ posting or receiving;
\item $\theta_{\tau }(\epsiQ,C)$ is the closeout cash flow at the first default $\tau = \tau_I \wedge\tau_C$, inclusive of the trading CVA and trading DVA cash flows after collateralization; also, we denote by $\bartau = T \wedge \tau$ the effective maturity of the contract;
\item $\varphi_{f,h}(t,s)$ are the discounted cash flows for the cost-of-funding-the-trade-accounts, representing flows of interest remuneration or cost due to the implementation of the trading strategy;
\item $\psi (t,\bartau , \tau_E )$ is the closeout cash flow for the external borrowing and lending activity the bank's treasury is doing to fund trading activities of the bank; we denote by $\tau_E$ the default time of the external lender/borrower.
\end{itemize}
 Our goal is to value the trade by taking to conditional expectation of discounted cash flows under the risk-neutral probability measure $\Q^r$, which is defined by postulating that the price of the (non-dividend-paying) risky asset $S$ discounted by the risk-free cash account $B^r$ is a local martingale. For concreteness, we will postulate in Section \ref{sec2.5} that the dynamics of $S$ under $\Q^r$ are
 \begin{align} \label{dynS}
 dS_t = S_t \big(r_t\,dt+\sigma \,dW^r_t\big)
 \end{align}
where $W^r$ is the Brownian motion under $\Q^r$. It is important to stress, however, that our approach covers a broad spectrum of semimartingale models, as will be demonstrated in Section \ref{sec3}. The filtration $\ff$ is generated by the Brownian motion $W^r$ and the full filtration $\gg$ is such that $\ff \subset \gg$ and $\tau_I$ and $\tau_C$ are $\gg$-stopping times. For simplicity, we will usually assume that $\Q^r (\tau_I = \tau_C)=0$. We will frequently use the shorthand notation $\Exx{t}{\cdot }$ for the conditional expectation $\mathbb{E}_{\Q^r}( \cdot \,|\, {\cal G}_t)$. Finally, for any real number $x$, we write $x = x^+ - x^-$ where $x^+ := x \vee 0$ and $x^- := (-x) \vee 0$.

\subsection{Valuation of Promised Cash Flows} \label{sec2.1}


Let the process $A$ represent all promised cash flows of the contract and let $\Pitoa(t,s,A)$ represent all payoffs from the contract between the dates $t$ and $s>t$, discounted at $t$ using the risk-free rate $r$
\begin{align*}
\Pitoa(t,s,A):= \int_{(t,s]} D_r(t,u) \, dA_u
\end{align*}
where $D_r(t,s) := B^r_t (B^{r}_s)^{-1} =  e^{-\int_t^s r_u \, du}$.
By the risk-neutral valuation for the uncollateralized defaultable contract $(A , \tau )$, we mean the equality
\begin{align} \label{eq2.1}
\Vtop^r_t(A,\tau ):=  \E_{\Q^r} \big[ \Pitoa(t,\bartau, \wt A)\,|\, \G_t \big] = \E^r_t \big[ \Pitoa(t,\bartau, \wt A) \big]
\end{align}
where we set $\wtA_t :=\I_{\{t < \tau \}} A_t+\I_{\{t \geq \tau \}} A_{\tau-}$ so that $\wt{A}$  gives the cash flows of $A$ that are stopped either just before the first default or at the maturity date $T$ if the first default happens after $T$. Note that $\Vtop^r_t(A,\tau )$ is simply the traditional risk-neutral price without collateralization and differential funding costs under the postulate of zero recovery at default, that is, with the null closeout payoff. Hence it should not be confused with the risk-neutral price $\pi^r_t(A):= \E^r_t [\Pitoa (t,T,A)]$ of an `equivalent' default-free version of the contract.

\subsection{Cost of Collateralization}         \label{sec2.2}


A second contribution to the price is due to the impact of the collateralization procedure, which is also known as `margining' among
practitioners. Let $C_t$ denote the level of the collateral account at time $t$, as specified by the contract's credit support annex (CSA) and let $\gamma(t,s,C)$ stand for the overall collateral margining costs between the times $t$ and $s$. By convention, $C_t>0$ means that the collateral has been overall posted by the counterparty to protect the trader, and the trader has to pay instantaneous interest $c^b_t$ on the related amount. If $C_t <0$, then the trader posts collateral for the counterparty and is remunerated at interest $c^l_t$. Therefore, the discounted net cash flows due to the margining procedure are given by
\[
\gamma(t,s,C) = \int_t^s D_r(t,u)C_{u}(r_u-\bar{c}_{u})\,du
\]
where the {\it effective collateral accrual rate} $\bar{c}$ is given by $\bar{c}_t := c^b_t \ind{C_t>0} + c^l_t \ind{C_t<0}$.
Note that if the collateral rates $c^b$ and $c^l$ are both equal to the risk-free rate $r$, then the cash flows $\gamma(t,s,C)$ vanish. By adding the margining costs to the price $\Vtop^r_t(A,\tau )$ given be \eqref{eq2.1}, we obtain the risk-neutral price for the collateralized defaultable contract $(A,C,\tau )$
\begin{align} \label{eq2.2}
\Vtop^r_t (A,C,\tau):=   \E^r_t \big[ \Pitoa(t,\bartau ,\wt{A}) + \gamma(t,\bartau,C) \big] = \Vtop^r_t(A,\tau )+ \LVA_t
\end{align}
where $\LVA_t := \Exx{t}{\gamma(t,\bartau,C)}$ is called the {\it collateral liquidity valuation adjustment}.

\subsection{Closeout Cash Flows}        \label{sec2.3}


As a third contribution, we consider the cash flow to/from the trader at the first default. One of the key financial aspects of the counterparty risky contract is the  credit support annex (CSA) closeout payoff, which occurs if at least one of the parties defaults either before or at the maturity of the contract. To define the CSA closeout payoff $\theta_{\tau }$ on the event $\{ \tau \leq T \}$, we first define the random variable $\Upsilon = Q_{\tau} - C_{\tau-}$ where $Q$ is the CSA {\it closeout valuation} process of the contract inclusive of the increment $\Delta A_\tau =A_\tau - A_{\tau-}$ representing a (possibly null) promised bullet dividend at $\tau$ and $C_{\tau-}$ is the value of the collateral process $C$ at the moment of the first default. Note that since the margin account is not updated at the moment of the first default, it is formally given by
$\wtC_t =\I_{\{t < \tau \}} C_t+\I_{\{t \geq \tau \}} C_{\tau-}$ so that $\wtC_{\tau } = \wtC_{\tau-}$.

In the financial interpretation, $\Upsilon^+$ is the amount the counterparty owes to the trader at time $\tau $, whereas $\Upsilon^-$ is the amount the trader owes to the counterparty at time $\tau $. It accounts for the legal value $Q_\tau$ of the contract, plus the bullet dividend $\Delta A_\tau$ to be received/paid at time $\tau$, less the collateral amount $C_{\tau-}$ since it is already held by either the trader (if $C_{\tau-}>0$) or the counterparty (if $C_{\tau-}<0$). We refer the reader to Section~3.1.3 in Cr\'epey et al. \cite{CBB2014} for more details regarding the financial interpretation of $\Upsilon$. The following definition describes the closeout payoff from the perspective of the trader. The random variables $R_I$ and $R_C$, which take values in $[0,1]$, represent the {\it recovery rates} of the trader and the counterparty, respectively.

In practice, the closeout cash flow $\theta_{\tau}(\epsiQ,C)$ can be calculated by following ISDA documentation. As customary, we define the cash flow $\theta_{\tau}(\epsiQ,C)$ by including the pre-default value of the collateral account used by the closeout netting rule to reduce exposure represented by the closeout valuation process $\epsiQ$ specified by the credit support annex. Here, we assume the collateral account can be re-hypothecated, see \cite{BCPP13} for a discussion on how collateralization impacts the closeout specification.

\bd \label{close}
{\rm The  {\it CSA closeout payoff} $\theta_{\tau }(\epsiQ,C) := \recR_{\tau } + C_{\tau -}$ on the event $\{\tau \leq T\}$ where the {\it recovery payoff} $\recR_{\tau }$ is given by the following expression}
\begin{equation} \label{closeout}
\recR_{\tau } := \I_{\{\tau_C< \tau_I \}}(R_C \Upsilon^+-\Upsilon^-) +
 \I_{\{\tau_I < \tau_C \}}(\Upsilon^+ - R_I \Upsilon^-) + \I_{\{\tau_I=\tau_C \}}( R_C \Upsilon^+ - R_I \Upsilon^-) .
\end{equation}
\ed

Let us comment on the form of the closeout payoff $\theta_{\tau }(Q,C)$. The term $C_{\tau -}$ reflects the fact that the legal title to the collateral amount comes into force only at the moment of the first default. The three terms in the right-hand side in \eqref{closeout} correspond to the CSA convention that, in principle, the net nominal cash flow at the first default, as seen from the perspective of the trader, should coincide with the closeout valuation $Q_{\tau}$ of the contract.

To identify various valuation adjustments caused by the counterparty credit risk, it is convenient to represent the net payoff at default as follows
\bde
\theta_{\tau }(\epsiQ,C) = Q_{\tau } + \I_{\{\tau_I < \tau_C \}} L_I \Upsilon^-  -  \I_{\{\tau_C< \tau_I \}} L_C \Upsilon^+ + \I_{\{\tau_I=\tau_C \}}(  L_I \Upsilon^- - L_C \Upsilon^+ )
\ede
where $L_C = 1- R_C$ (respectively, $L_I=1-R_I$) is the {\it loss given default} coefficient for the counterparty (respectively, the trader). Obviously, the loss given default coefficient equal to one (respectively, zero) corresponds to the case of null (respectively, full) recovery coefficient. When $L_C = L_I =0$, then we obtain $\theta_{\tau }(\epsiQ,C) = Q_{\tau }$, so indeed we deal here with the full recovery of the CSA closeout value of the contract. However, due to the presence of collateral, in the case of the total loss given default, that is, when $L_C = L_I =0$, the CSA closeout payoff to the bank equals
\begin{align*}
\theta_{\tau }(\epsiQ,C) &= \I_{\{\tau_C< \tau_I \}} \big(Q_{\tau } \I_{\{  Q_{\tau } < C_{\tau-}\}} + C_{\tau-} \I_{\{ Q_{\tau } \geq C_{\tau-} \}} \big) + \I_{\{\tau_I< \tau_C \}} \big( Q_{\tau } \I_{\{  Q_{\tau } \geq C_{\tau-} \}} + C_{\tau-} \I_{\{ Q_{\tau } < C_{\tau-}    \}} \big)  \\
&+ \I_{\{\tau_C = \tau_I \}}  C_{\tau-}
\end{align*}
and thus the full recovery of the CSA closeout value for the trader or the counterparty may still occur in some circumstances.
For simplicity, we will henceforth assume that the event $\{\tau_C = \tau_I\}$ is negligible under $\Q^r$, so that
\be \label{deftt}
\theta_{\tau }(\epsiQ,C) = Q_{\tau } + \I_{\{\tau = \tau_I \}} \lgd_I \Upsilon^- -  \I_{\{\tau = \tau_C \}} \lgd_C \Upsilon^+
\ee
or, equivalently,
\begin{align}  \label{cc1}
\theta_{\tau}(\epsiQ,C):= \epsiQ_{\tau}+\ind{\tau=\tau_I}\Pitoa_{\mbox{\footnotesize DVA}}-\ind{\tau=\tau_C}\Pitoa_{\mbox{\footnotesize CVA}}
\end{align}
where we denote $\Pitoa_{\mbox{\footnotesize DVA}}=\lgd_I(\epsiQ_{\tau} - C_{\tau-})^-= \lgd_I \Upsilon^-$ and
$\Pitoa_{\mbox{\footnotesize CVA}}=\lgd_C(\epsiQ_{\tau} - C_{\tau-})^+ =\lgd_C \Upsilon^+$.
After accounting for the on-default cash flow, we obtain the following representation for the risk-neutral price of the collateralized defaultable contract $(A,C,R)$
\begin{align} \label{eq2.3}
\Vtop^r_t(A,C,R,\tau ) := \E^r_t \big[ \Pitoa(t,\bartau ,\wt{A})+\gamma(t,\bartau,C)+ \vartheta (t, \tau ,\epsiQ ,C) \big]
\end{align}
where $\vartheta (t, \tau ,\epsiQ , C) := \ind{\tau \leq T}D_r(t,\tau)\theta_\tau(\epsiQ,C)$. As customary, we define the {\it debit valuation adjustment} $\DVA_t$
\[
\DVA_t = \E^r_t \big[\ind{\tau = \tau_I \leq T} D_r(t,\tau ) \lgd_I(\epsiQ_{\tau} - C_{\tau-})^- \big]
\]
and  the {\it credit valuation adjustment} $\CVA_t$ by
\[
\CVA_t = \E^r_t \big[\ind{\tau = \tau_C \leq T} D_r(t,\tau ) \lgd_C(\epsiQ_{\tau} - C_{\tau-})^+ \big].
\]
Then the risk-neutral price for the collateralized defaultable contract $(A,C,R,\tau )$ with recovery $R$ can be represented as follows
\begin{align} \label{eq2.2n}
\Vtop^r_t (A,C,R,\tau) = \Vtop^r_t(A,\tau )+ \LVA_t + \E^r_t \big[ \ind{\tau \leq T} \epsiQ_{\tau} \big] + \DVA_t - \CVA_t .
\end{align}
To proceed further, we need to specify the closeout valuation process $\epsiQ$.
For instance, under the risk-free closeout we have that $\epsiQ_\tau := \Exx{\tau}{ \Pitoa(\tau,T)}$, whereas if the replacement
closeout is postulated, then we set $\epsiQ_\tau :={\Vtop}^r_\tau(A,C,\recR ,\tau )$ where the value $\Vtop^r_t(A,C,R ,\tau )$ is given by \eqref{eq2.3}. Note that in the latter case we obtain a nonlinear recursive equation for the contract's risk-neutral price (see Brigo and Morini \cite{BM2010} or Durand and Rutkowski \cite{DR2013} for more details).

\subsection{Funding Costs and Benefits}    \label{sec2.4}


In this step, we focus on the funding costs of the hedging strategy and we add the relevant cash flows by adopting the procedure proposed in Pallavicini et al. \cite{PPB12}. Let $F_t$ be the cash account for the replication of the trade and let $H_t$ stand for the value of the trader's positions in the risky asset $S$. We work under the repo trading convention, meaning that the risky asset
$S$ is funded using a cash account $F^S_t$ and the equality $F^S_t=- H_t$ holds for every $t\in [0,T]$. The case of collateralized risky assets can be treated in the same way by interpreting the cash account as the collateral account for such assets. We denote by $\varphi_{f,h}(t,s) = \varphi_f(t,s)+ \varphi_h(t,s)$ the discounted incremental funding costs between $t$ and $s$. If we remove the tacit assumption that the trader can borrow cash at the risk-free rate, we obtain the following cash flows for either the cost/benefit of carry of the cash account
\begin{align*}
\varphi_f(t,s) := \int_t^s D_r(t,u)F_{u}(\bar{f}_{u}-r_u)\,du
\end{align*}
and for either the cost/benefit of carry of the repo transaction
\begin{align*}
\varphi_h(t,s) := \int_t^s D_r(t,u)F^S_{u}(\bar{h}_u-r_u)\,du
\end{align*}
where the {\it effective funding rate} $\bar{f}$ equals $\bar{f}_t := f^l_t \ind{F_t \geq 0}+f^b_t \ind{F_t<0}$
and the {\it effective repo rate} $\bar{h}$ is given by $\bar{h}_t := h^l_t \ind{F^S_t\geq 0}+h^b_t \ind{F^S_t<0}$.
If we distinguish borrowing and lending of cash from treasury explicitly, then we obtain
\begin{align*}
\varphi_f(t,s)=\int_t^s D_r(t,u)\big(F_u^+(f^l_u-r_u)-F_u^-(f^b_u-r_u)\big)\,du = \varphi_{f^l}(t,s)-\varphi_{f^b}(t,s).
\end{align*}
If the trader can borrow and lend at risk-free rate so that $\bar{f} = r$, then $\varphi_f(t,s)$ vanishes.
A similar analysis applies to the term $\varphi_h(t,s)$ and we thus get the following decomposition
\begin{align*}
\varphi_h(t,s)=\int_t^s D_r(t,u)\big( (F^S_u)^+(h^l_u-r_u)- (F^S_u)^-(h^b_u-r_u)\big)\,du = \varphi_{h^l}(t,s)-\varphi_{h^b}(t,s).
\end{align*}
The risk-neutral price of the collateralized defaultable contract $(A,C,R)$ inclusive of funding costs is given by
\begin{align} \label{eq2.4}
\Vtop^{r,f,h}_t (A,C,R,\tau )&:=\E^r_t \big[ \Pitoa(t,\bartau,\wt{A})+\gamma(t,\bartau,C)+ \vartheta (t, \tau ,\epsiQ ,C)+\varphi_{f,h}(t,\bartau )\big] \nonumber \\  &= \Vtop^r_t(A,\tau )+ \LVA_t+ \E^r_t \big[ \ind{\tau \leq T} \epsiQ_{\tau} \big] + \DVA_t - \CVA_t + \FVA_t
\end{align}
where the term $\FVA_t := \Exx{t}{\varphi_{f,h}(t,\bartau )}$ represents the {\it funding valuation adjustment}.
When $f^b \geq f^l \geq r$ and $h^b \geq h^l \geq r$, the {\it funding benefit adjustments} are equal to
\[
\FBA^f_t = \Exx{t}{\varphi_{f^l}(t,s)}, \quad \FBA^h_t = \Exx{t}{\varphi_{h^l}(t,s)}
\]
and the {\it funding cost adjustments} are given by
\[
\FCA^f_t = \Exx{t}{\varphi_{f^b}(t,s)}, \quad \FCA^h_t = \Exx{t}{\varphi_{h^b}(t,s)}.
\]
Obviously, we have that $\FVA_t = \FBA^f_t + \FBA^h_t - (\FCA^f_t + \FCA^h_t)$. The funding policy of the bank's treasury is determined by funding rates for cost, $f^b$, and  benefit, $f^l$, of carry of hedge accounts, which both depend on the funding policy of the bank.

\subsection{Invariance Property}           \label{sec2.5}

For brevity, we will sometimes write $\Vtop^{r,f,h}_t$ instead of $\Vtop^{r,f,h}_t(A,C,R,\tau )$, which is given by \eqref{eq2.4}. It is puzzling that the price $\Vtop^{r,f,h}_t$ exhibits an apparent dependence on the non-accessible risk-free interest rate $r$ through the discount factor $D_r(t,s)$ and the presence of the risk-free rate $r_t$ in dynamics \eqref{dynS} of the risky asset $S$ under $\Q^r$.  As in Pallavicini et al. \cite{PPB12}, we will show that this dependence is illusory, by deriving an equivalent representation for the price in which the risk-free rate does not appear at all. This means that the price of a contract is in fact invariant with respect to the specification of the risk-free rate $r_t$. In Proposition \ref{pro2.1}, we will show that the invariance property is valid within the setup studied in this section. In Section \ref{sec3}, we will prove, using different arguments, that this crucial feature is valid in a general semimartingale framework as well.

Recall that the filtration $\ff$ is generated by the Brownian motion $W^r$. For the purpose of Section \ref{sec2.5}, we postulate that the default times $\tau_C$ and $\tau_I$ are $\ff$-conditionally independent and have the $\ff$-hazard rates $\lambda^C$ and $\lambda^I$, respectively. Then the so-called {\it immersion property} between $\ff$ and $\gg$ holds under $\Q^r$. If, in addition, the process $A$ is $\ff$-adapted, then by means of credit risk computations we obtain the following equality, which holds on the event $\{t< \tau \}$ for every $t \in [0,T]$,
\begin{align*}
\Vtop^{r,f,h}_t =\E^r \bigg( \int_t^T D_{r+\lambda }(t,u)\Big( \big( ( r_u-\bar{c}_u)C_u+(\bar{f}_u-r_u)F_u
+(\bar{h}_u-r_u)F^S_u+\lambda_u\theta_u \big)\,du+dA_u \Big)\,\Big|\,{\cal F}_t\bigg)
\end{align*}
where $\lambda_t:=\lambda^C_t+\lambda^I_t$ is the hazard rate of the first default and, for all $t \leq s$,
\[
D_{r+\lambda }(t,s):=e^{-\int_t^s (r_u+\lambda_u)\,du} = B^{r+\lambda}_t (B^{r+\lambda}_s)^{-1}.
\]
Let $\Q^h$ be a probability measure on $(\Omega , {\cal G}_T)$ such that the dynamics of the process $S$ under $\Q^h$ are
\[
dS_t=S_t\big(\bar{h}_t\,dt+\sigma\,dW^h_t\big)
\]
where $W^h$ is a Brownian motion under $\Q^h$ (the notation $\Q^{\bar{h}}$ would be more appropriate, but also too cumbersome).
From the Girsanov theorem, it is known that $dW^h_t=dW^r_t-\sigma^{-1}(\bar{h}_t-r_t)\,dt$.
We set, for all $t \leq s$,
\[
D_{\bar{f}+\lambda}(t,s):=e^{-\int_t^s(\bar{f}_u+\lambda_u)\,du} = B^{\bar{f}+\lambda}_t (B^{\bar{f}+\lambda}_s)^{-1}.
\]

\bp \label{pro2.1}
Assume that $\pi_t = v(S_t,t)$ where $v \in C^{2,1}(\rr_+ \times [0,T],\rr )$.
Then the equality $\Vtop^{r,f,h}_t (A,C,R,\tau )= \Vtop^{f,h}_t(A,C,R,\tau )$ holds on the event $\{t< \tau \}$ for every $t \in [0,T]$
where
\begin{align}  \label{xx1}
\Vtop^{f,h}_t(A,C,R,\tau ) :=\E^{h}\bigg(\int_t^T D_{\bar{f}+\lambda}(t,u)\Big( \big( (\bar{f}_u-\bar{c}_u)C_u+\lambda_u\theta_u \big) du+dA_u\Big)\,\Big|\,{\cal F}_t \bigg)
\end{align}
so that, for every $t \in [0,T]$,
\begin{align} \label{eq:totawsV}
\Vtop^{f,h}_t(A,C,R,\tau ) =  \E^h_t \big[ \Pi_{\bar{f}}(t,\bartau ,\wt{A})+\gamma_{\bar{f}}(t,\bartau,C)+\vartheta_{\bar{f}}(t, \tau ,\epsiQ, C)  \big]
\end{align}
where the conditional expectation $\mathbb{E}^h_t [ \,\cdot \,]:=\mathbb{E}_{\Q^h} [ \, \cdot \,|\, {\cal G}_t]$ is computed under $\Q^h$ and where we denote
\begin{align*}
&\Pi_{\bar{f}}(t,\bartau ,\wt{A}):= \int_{(t,\bartau ]} D_{\bar{f}}(t,u) \, d\wt{A}_u ,\\
&\gamma_{\bar{f}}(t,\bartau ,C) := \int_t^{\bartau } D_{\bar{f}}(t,u)C_{u}( \bar{f}_u-\bar{c}_{u})\,du , \\
&\vartheta_{\bar{f}}(t, \tau ,\epsiQ , C) := \ind{\tau \leq T}D_{\bar {f}}(t,\tau)\theta_\tau(\epsiQ,C).
\end{align*}
\ep

\proof
Let us further simplify the notation by writing $\pi_t = \Vtop^{r,f,h}_t$. Let $V^p_t$ be the value at time $t$ of the replicating portfolio (see Section~\ref{sec3} for more details on the self-financing condition under funding costs and the concept of replication under collateralization).
Then we have  $V^p_t = F_t$ and $-\pi_t = V^p_t - C_t =F_t-C_t$ where the minus sign means that we consider here the price the trader is willing
to `pay' for the contract. Consequently, for every $t \in [0,T]$,
\begin{align*}
\pi_t  =\E^r \bigg( \int_t^T D_{r+\lambda }(t,u)\Big( \big( (\bar{f}_u-\bar{c}_u)C_u + (r_u-\bar{f}_u)\pi_u
+(\bar{h}_u-r_u)F^S_u+\lambda_u\theta_u \big) du+dA_u \Big)\,\Big|\,{\cal F}_t\bigg).
 \end{align*}
Observe that the process $\wt \pi_t := \pi_t (B_t^{r+\lambda })^{-1}$ satisfies
 \begin{align*}
\wt{\pi_t}=\, &\E^r \bigg( \int_0^T \Big( \big( (\bar{f}_u-\bar{c}_u)\wt{C}_u + (r_u-\bar{f}_u)\wt{\pi}_u
+(\bar{h}_u-r_u)\wt{F}^S_u+ \lambda_u \wt{\theta}_u \big) du+ (B^{r+\lambda }_u)^{-1} dA_u \Big)\,\Big|\,{\cal F}_t\bigg) \\
& - \int_0^t \Big( \big( (\bar{f}_u-\bar{c}_u)\wt{C}_u + (r_u-\bar{f}_u) \wt{\pi}_u
+(\bar{h}_u-r_u) \wt{F}^S_u+\lambda_u \wt{\theta}_u \big)\,du+ (B^{r+\lambda }_u)^{-1} dA_u \Big)
 \end{align*}
 where $\wt{C}_t :=  C_t (B_t^{r+\lambda })^{-1}$ and $\wt{\theta}_t := \theta_t (B_t^{r+\lambda })^{-1}$.
Under the assumption that $\pi_t = v(S_t,t)$, using the It\^o formula, we obtain under $\Q^r$
\begin{align*}
d\wt{\pi}_t =\, & \big( (\bar{c}_t- \bar{f}_t ) \wt{C}_t+(\bar{f}_t-r_t)\wt{\pi}_t
+(r_t-\bar{h}_t)\wt{F}^S_t+\lambda_u \wt{\theta}_t \big) dt \\ & - (B_t^{r+\lambda })^{-1}\, dA_t
+  (B_t^{r+\lambda })^{-1} \frac{\partial {v}}{\partial s}(S_t,t)\, S_t \sigma \, dW^r_t .
\end{align*}
Using again the It\^o formula, we deduce that the process $\wh \pi_t:=(B_t^{\bar{f}+\lambda })^{-1}\pi_t=\wt{\pi}_t B^r_t (B^{\bar f}_t)^{-1}$
satisfies
\[
d\wh \pi_t:=(r_t-\bar{f}_t)\wh{\pi}_t+B^r_t(B^{\bar f}_t)^{-1}\,d\wt{\pi}_t
\]
and thus
\[
d\wh{\pi}_t=\big((\bar{c}_t-\bar{f}_t)\wh{C}_t+(\bar{h}_t -r_t)\wh{F}^S_t +\lambda_t \wh{\theta}_t\big) dt
-(B_t^{\bar{f}+\lambda})^{-1}\,dA_t+(B_t^{\bar{f}+\lambda })^{-1}\frac{\partial{v}}{\partial s}(S_t,t)\,S_t\sigma\,dW^r_t
\]
where $\wh{C}_t :=  C_t (B_t^{\bar{f}+\lambda })^{-1}$ and $\wh{\theta}_t := \theta_t (B_t^{\bar{f}+\lambda })^{-1}$. This also means that
\begin{align*}
\pi_t =\E^r \bigg( \int_t^T D_{\bar{f}+\lambda}(t,u)\Big( \big( (\bar{f}_u-\bar{c}_u)C_u
+(r_u-\bar{h}_u )F^S_u+\lambda_u\theta_u \big)\,du+dA_u \Big)\,\Big|\,{\cal F}_t\bigg).
 \end{align*}
To eliminate $F^S$ from the formula above, we observe that the hedge ratio satisfies
\[
 H_t = S_t \frac{\partial {v}}{\partial s}(S_t,t) = - F^S_t .
\]
Consequently, since $dW^h_t=dW^r_t-\sigma^{-1}(\bar{h}_t-r_t)\,dt$, we see that  the process $\wh \pi_t $ satisfies under $\Q^h$
\[
d\wh{\pi}_t = \big( (\bar{c}_t- \bar{f}_t ) \wh{C}_t +\lambda_t \wh{\theta}_t \big)\,dt - (B_t^{\bar{f}+\lambda})^{-1}\, dA_t + (B_t^{\bar{f}+\lambda })^{-1} \frac{\partial {v}}{\partial s}(S_t,t)\, S_t \sigma \, dW^h_t .
\]
Note that the immersion property between $\ff$ and $\gg$ still holds under $\Q^h$. We thus conclude that on the event $\{t< \tau \}$ for every $t \in [0,T]$
\begin{align*}
\pi_t =\E^h \bigg( \int_t^T D_{ \bar{f} +\lambda}(t,u)\Big( \big( (\bar{f}_u-\bar{c}_u)C_u
 +\lambda_u\theta_u \big)\,du+dA_u \Big)\,\Big|\,{\cal F}_t\bigg) =: \Vtop^{f,h}_t(A,C,R,\tau ).
 \end{align*}
Finally, it is easy to see that equality \eqref{eq:totawsV} is valid as well.
\endproof

Proposition \ref{pro2.1} will be generalized in Section \ref{sec3} to a general semimartingale measure (see Corollary \ref{cor1.1}).
Note, however, that the convention for the trader's price in Sections \ref{sec2} and \ref{sec3} is slightly different. To be more specific, if $p$ stands for the initial price of a contract, then the initial value of the trader's portfolio equals $-p+C_0$ in Section \ref{sec2} whereas it equals $p+C_0$ in Section \ref{sec3}.

Recall that $\E^h$ is the expected value under a probability measure where the underlying asset has the rate of return $\bar{h}$.
Since $\bar{h}$ depends on $H$ (and hence on $\pi^{f,h}$), the pricing measure depends on the future values of the very price we
are computing, meaning that the pricing measure becomes deal dependent.

\subsection{Credit and Funding Valuation Adjustments}    \label{sec2.6}

The next lemma gives a formal representation of the price $\Vtop^{f,h}_t$ as the risk-neutral price
of an equivalent non-defaultable contract complemented by the valuation adjustments. Let us denote $\Delta A_{\tau }=
A_{\tau } - A_{\tau -}$.

\bl \label{lemm2.1}
{\it Assume the risk-free closeout valuation $\epsiQ_\tau = \Delta A_{\tau } + \Exx{\tau}{ \Pitoa(\tau,T,A)} = \Delta A_{\tau } + \pi^r_{\tau}(A)$  where $\pi^r_t(A):= \E^r_t [\Pitoa (t,T,A)]$ is the risk-neutral price of the equivalent non-defaultable version of the contract. Then} \rm
\begin{align*} 
\Vtop^{f,h}_t(A,C,R,\tau ) = \pi^r_t(A)+ \LVA_t + \DVA_t - \CVA_t + \FVA_t .
\end{align*}
\el

\proof
The assertion is an easy consequence of \eqref{eq2.2}, \eqref{cc1} and \eqref{eq2.4}.
\endproof

Recall that
\[
\Vtop^{f,h}_t(A,C,R,\tau ) = \E^r_t \big[ \Pitoa(t,\bartau ,\wt{A})+\gamma(t,\bartau,C)
   + \vartheta (t, \tau ,\epsiQ ,C) +\varphi_{f,h}(t,\bartau) \big].
\]
The following question thus arises in the context of Lemma \ref{lemm2.1}: can we interpret the conditional expectation
$\E^r_t \big[ \Pitoa(t,\bartau ,\wt{A})+ \vartheta (t, \tau ,\epsiQ ,C) \big]$ as $\pi^r_t(A) + \DVA_t - \CVA_t$ and, separately, the conditional expectation $\Exx{t}{\gamma(t,\bartau,C)+\varphi_{f,h}(t,\bartau)}$
as $\LVA_t + \FVA_t $? Not really, since in fact we deal here with a non-linear equation in which, for instance,
the present value of $\varphi_{f,h}(t,\bartau )$ depends via $\bar{f}$ and $\bar{h}$ of on future values
of $F_s$ and $F^S_s$ for $s \geq t$, and thus it depends on future values of $\Vtop^{f,h}_t$.
In fact, all terms feed each other and there is no neat separation of risks. For this reason, we will analyze in Section \ref{sec3} pricing through nonlinear BSDEs.

\subsection{External Funding Adjustments}  \label{sec2.7}



In Section \ref{sec2.7}, we extend our simple model of the bank by including also the external lender/borrower.
We denote by $\psi$ the cash flows the bank has in place with the external lender/borrower to service the
trade, inclusive of closeout if the external entity or the bank's default. In principle, the final price of a new contract should also reflect its impact on the cash flows from debit and credit risk in the funding strategy of the bank. We are going to consider
two alternative funding strategies of the bank relative to the bank's treasury and the external entity.

Let $Y_t$ be the bank's borrowing/lending at time $t$ associated with all existing trades, but excluding the new contract we are valuing. By convention, $Y_t<0$ means that the bank borrows from the external entity and $Y_t>0$ corresponds to lending. Then the corresponding cash flows at default associated with the external entity are given by the following expression
\begin{align} \label{efa11}
\Psi (X) := \I_{\{\tau \wedge \tauE = \tau_I \leq T\}} \lgd_I Y_{\tau}^-
 -  \I_{\{\tau \wedge \tauE = \tauE \leq T \}}  \lgdE Y_{\tauE}^+
\end{align}
where $\tauE$ is the default time of the {\it external entity} and where we assume that the event $\{\tau = \tauE\}$ is negligible under $\Q^r$. Obviously, the incremental impact due to the external funding of the new trade on these cash flows is given by $\Psi (Y+F) - \Psi (Y)$. More explicitly, the discounted cash flows are given by
\begin{align} \label{efa12}
\psi(t,\bartau , \tauE ) &=  D_r(t,\tau) \I_{\{\tau \wedge \tauE = \tau_I \leq T\}} \lgd_I \big( (Y_{\tau }+ F_{\tau})^- - Y_{\tau}^- \big)
\\& - D_r(t,\tauE) \I_{\{\tau \wedge \tauE = \tauE \leq T \}}  \lgdE \big( (Y_{\tauE}+F_{\tauE})^+ - Y_{\tauE}^+\big). \nonumber
\end{align}

Suppose first that the treasury funding of the new trade is done independently of other trading activities of the bank. In that case, we may set $Y = 0$ in \eqref{efa12}, so that it becomes
\begin{align} \label{efa13}
\psi(t,\bartau , \tauE ) = D_r(t,\tau) \I_{\{\tau \wedge \tauE = \tau_I \leq T\}} \lgd_I F_\tau^-
 - D_r(t,\tauE) \I_{\{\tau \wedge \tauE = \tauE \leq T \}}  \lgdE F_{\tauE}^+
\end{align}
and thus
\begin{align*}
\E^r_t[\psi (t,\bartau , \tauE )]& = \E^r_t \big[ D_r(t,\tau) \I_{\{\tau \wedge \tauE = \tau_I \leq T\}} \lgd_I F_{\tau}^- \big]
 - \E^r_t \big[ D_r(t,\tauE) \I_{\{\tau \wedge \tauE = \tauE \leq T \}}  \lgdE F_{\tauE}^+ \big] \\
 &= \DVAF_t - \CVAF_t.
\end{align*}
The debit valuation adjustment due to external borrowing, denoted as $\DVAF$, is also occasionally called the {\it funding debit adjustment} (FDA) or DVA$_2$ in the existing literature. Its presence is due to the fact that the bank profits from its own default by not paying back its debt in full to the external lender. Hence this term corresponds to the CVA on the loan that the external lender will charge the bank upon entering a loan. Let us stress that under the adopted external funding convention the term $\DVAF_t$ is always nonnegative, which can be here explained by the benefit at the bank's default when the trading desk is borrowing from the treasury at the moment of the bank's default so that $F_{\tau}^- >0$. Similarly, the term $-\CVAF_t$ is the credit valuation adjustment triggered when the bank's treasury is lending to the external entity and the borrower defaults first.

We henceforth focus on an alternative (and practically important) external funding convention where the treasury funding across all trades is netted and the bank is assumed to always be a net borrower. In that case, the incremental amount $F_t^-$ associated with a new contract is not used by the bank's treasury to start an external loan, but rather to reduce the overall bank's borrowing.  Formally, we assume that $Y_t \leq 0$ and $Y_t + F_t \leq 0 $ for every $t \in [0,T]$. Then  \eqref{efa12} yields
\begin{align} \label{efa2}
\psi(t,\bartau) := - D_r(t,\tau)\I_{\{\tau=\tau_I \leq T\}}\lgd_I F_\tau =D_r(t,\tau)\I_{\{\tau=\tau_I \leq T\}}\lgd_I(F_\tau^- - F_\tau^+)
\end{align}
where we assumed, without loss of generality, that the external entity is non-defaultable. We set
\[
\E^r_t [\psi(t,\bartau )] = \E^r_t \big[ D_r(t,\tau) \I_{\{\tau = \tau_I \leq T\}} \lgd_I (F_\tau^- - F_\tau^+) \big]
 = \DVAF_t = \DVAFn_t - \DVAFp_t
\]
where, manifestly, the debit valuation adjustment $\DVAF_t$ can be either positive or negative. The latter case corresponds to the situation where a new trade may reduce the benefit at the bank's default by reducing the overall bank's debt (this occurs, in particular, when $F_{\tau }>0$).

To examine in some detail the impact of the external funding on the price of a contract, let us consider a contract with a single payoff $X$ at time $T$ so that $A_t = X \I_{\{t=T\}}$. We wish to compute the price for $X$ inclusive of external funding adjustments under the risk-free closeout valuation $\epsiQ_\tau = \Exx{\tau}{ \Pitoa(\tau,T)} = \E^r_{\tau } [D_r(\tau ,T)X ] =: \pi^r_{\tau }(X)$. For brevity, we assume that $h^b=h^l=r$. Using Lemma \ref{lemm2.1}, we get the representation inclusive of the external funding benefits/losses.

\bp \label{profin}    
{\it Assume that $\epsiQ_\tau = \Delta A_{\tau } + \pi^r_{\tau}(X)$ and $h^b=h^l=r$. If $X \leq 0$ and $X+F \leq 0$, then} \rm
\begin{align} \label{repxx}
 &\Vtop^{f,h}_t (A,C,R,\tau ) = \pi^r_t(X)+ \LVA_t + \DVA_t - \CVA_t + \FBA^f_t - \FCA^f_t + \DVAFn_t - \DVAFp_t  \\
 & = \pi^r_t(X) + \E^r_t \bigg[ \int_t^{\bartau } D_r(t,u)C_{u}(r_u-\bar{c}_{u})\,du \bigg] + \E^r_t \big[  D_r(t,\tau)
 (\ind{\tau = \tau_I \leq T} \lgd_I\epsiQ_{\tau}^- - \ind{\tau = \tau_C \leq T}  \lgd_C \epsiQ_{\tau}^+ ) \big]  \nonumber
\\ &+ \E^r_t \bigg[ \int_t^{\bartau } D_r(t,u)\big(F_u^+(f^l_u-r_u)-F_u^-(f^b_u-r_u)\big)\,du \bigg]
 + \E^r_t \big[ D_r(t,\tau) \I_{\{\tau = \tau_I \leq T\}} \lgd_I (F_\tau^- - F_\tau^+) \big]. \nonumber
\end{align}
\ep

We will now examine two particular instances of Proposition \ref{profin} with $C=0$. Suppose first that the final payoff is nonnegative for the trader so that $X \geq 0$. Then $F_t \leq 0$ and $Q_t \geq 0$ for every $t \in [0,T]$.  Since $F^+_t=Q^-_t=0$,  \eqref{repxx} simplifies to
\begin{align*}
 \Vtop^{f,h}_t (A,R,\tau ) &= \pi^r_t(X) - \CVA_t - \FCA^f_t + \DVAFn_t  = \E^r_t [ D_r(t,T)X ]  - \E^r_t \big[ D_r(t,\tau) \ind{\tau = \tau_C \leq T}
 \lgd_C \epsiQ_{\tau}^+ \big] \\ & - \E^r_t \bigg[ \int_t^{\bartau } D_r(t,u) F_u^- (f^b_u-r_u)\,du \bigg]
 + \E^r_t \big[ D_r(t,\tau) \I_{\{\tau = \tau_I \leq T\}} \lgd_I F_\tau^-  \big].
\end{align*}
Note that the last two terms have the opposite effect on the price and, ideally, they may even cancel each other
if the equality $\FCA^f_t = \DVAFn_t$ occurs and thus the {\it net funding/default benefit} $\FCA^f_t-\DVAFn_t$ vanishes.

If, on the contrary, the final payoff to the trader is nonpositive so that $X \leq 0$, then $F_t \geq 0$ and $Q_t \leq 0$ so that $F^-_t=Q^+_t=0$. Consequently, \eqref{repxx} becomes
\begin{align*}
\Vtop^{f,h}_t (A,R,\tau ) &= \pi^r_t(X) + \DVA_t + \FBA^f_t - \DVAFp_t  = \E^r_t [ D_r(t,T)X ]  + \E^r_t \big[  D_r(t,\tau) \ind{\tau = \tau_I \leq T} \lgd_I \epsiQ_{\tau}^- \big] \\ & + \E^r_t \bigg[ \int_t^{\bartau } D_r(t,u) F_u^+ (f^l_u-r_u)\,du \bigg]
 - \E^r_t \big[ D_r(t,\tau) \I_{\{\tau = \tau_I \leq T\}} \lgd_I F_\tau^+  \big].
\end{align*}
If, in addition, $f^l_t = r_t$ for all $t \in [0,T]$, which is a practically plausible assumption, then it is not hard to obtain the equalities $\DVA_t =\DVAFp_t$ and $\Vtop^{f,h}_t = \pi^r_t(X)$ by arguing that $\epsiQ_{\tau } = - F_{\tau }$.
In that case, the valuation adjustments cancel each other and thus $\Vtop^{f,h}_t (A,R,\tau ) = \pi^r_t(X)$.

We conclude that the double counting of debit valuation adjustments does not appear since lending to the external lender, which reduces the bank's benefit at its default at the expense of the external borrower, is combined with the negative exposure at the bank's default, which in turn increases the bank's benefit at its default at the expense of the counterparty in the trade).

\section{Valuation in a General Semimartingale Model}   \label{sec3}

Throughout Section \ref{sec3}, we fix a finite trading horizon date $T>0$ for our model of the financial market given on a filtered probability space $(\Omega, {\cal G} , {\mathbb G}, \P)$ where the filtration  ${\mathbb G} = ({\cal G}_t)_{t \in [0,T]}$ satisfies the usual conditions of right-continuity and completeness. For convenience, we assume that the initial $\sigma$-field ${\cal G}_0$ is trivial. Moreover, all processes introduced in what follows are implicitly assumed to be $\mathbb G$-adapted and, as usual, any semimartingale is assumed to be c\`adl\`ag. 
Let us introduce the notation for interest rates and prices of all traded assets in our market model.

\noindent {\bf Treasury rates.} The \textit{lending} (respectively,  {\it borrowing}) \textit{cash account} $B^l$ (respectively,  $B^b$) can be used by the trader for unsecured lending (respectively, borrowing) of cash from the bank's treasury.
When the borrowing and lending treasury rates are equal,  the single treasury account is denoted by $\Bf$.
It is assumed that $d\Blr_t = \rll_t \Blr_t \, dt,\, d\Bbr_t = \rbb_t \Bbr_t \, dt$ and $d\Bf_t = \rlb_t \Bf_t \, dt $ where the
treasury funding rates $\rll , \rbb$ and $\rlb$ are $\gg$-adapted.

\noindent {\bf  Non-defaultable risky assets traded on repo market.}  We denote by $(S^1,S^2,\ldots,S^d)$ the collection of prices of $d$ risky assets, which do not pay dividends.   We denote by $\Bilr$ (respectively, $\Bibr$) the {\it lending} (respectively, {\it borrowing}) {\it repo account} corresponding to the $i$th risky non-defaultable asset.  In the special case when $\Bilr=\Bibr$, the single repo account is denoted by $B^i.$  Furthermore, we assume that $d\Bilr_t = \rill_t \Bilr_t \, dt,\, d\Bibr_t = \ribb_t \Bibr_t \, dt$ and $dB^i_t = \rilb_t B^i_t \, dt $ and the processes $S^1, S^2, \dots, S^d$ are $\gg$-semimartingales.

\noindent {\bf  Non-defaultable risky assets traded through treasury funding.}  Let $(S^{d+1},S^{d+2},\ldots,S^m)$ be the collection of prices of $m$ non-defaultable risky assets, which do not pay dividends. We assume that the processes $S^{d+1},S^{d+2}, \dots, S^{d+m}$ are $\gg$-semimartingales.

\noindent {\bf Defaultable bonds.} Let $D^1(t,T)$ be $D^2(t,T)$ are bonds issued by the trader's bank and the counterparty's entity.
Let $\tau = \tau_1 \wedge \tau_2 = \tau_I \wedge \tau_C $ where $\tau_1=\tau_I$ and $\tau_2 = \tau_C$ are  $\gg$-stopping times
representing the {\it default times} of the trader and the counterparty, respectively. We denote by $\bartau := \tau \wedge T$ the {\it effective maturity} of the contract.

\subsection{Linear Model with Funding Costs and Defaults}   \lab{sec3.1}

We start by examining a special case of the model with funding costs and defaults.
By the {\it linear setup}, we mean a particular instance of a general semimartingale model where:
\begin{itemize}
\item  risky assets $S^1,S^2,\dots ,S^d$ are traded on repo market with the corresponding repo accounts
$B^i$ for $i=1,2, \dots ,d$ and the associated repo rates $\rilb$,
\item risky assets $S^{d+1}, S^{d+2}, \dots, S^m$ and defaultable bonds $D^1(t,T)$ and $D^2(t,T)$  are traded with funding
through the bank's treasury account $\Bf$ with the treasury funding rate $\rlb$.
\end{itemize}

\subsubsection{Non-defaultable Uncollateralized  Contracts}   \lab{sec3.1.1}

The {\it value process} of a portfolio $\phi = (\psi^f, \kappa^1, \kappa^2, \psi^1 ,\dots, \psi^d, \xi^1, \dots , \xi^{m} )$ of traded assets and the corresponding funding accounts equals
\begin{align} \label{sef0}
V^p_t (\phi , A) & :=  \psif_t \Bf_t + \sum_{j=1}^2 \pdj_tD^j(t,T) +
\sum_{i=1}^d (\psi^i_t B^i_t + \xi^i_t S^i_t )  + \sum_{i=d+1}^m \xi^i_t S^i_t   \\
& = \psif_t \Bf_t+  \sum_{j=1}^2 \pdj_tD^j(t,T) + \sum_{i=d+1}^m \xi^i_t S^i_t = F_t  + \sum_{j=1}^2 D^j_t + \sum_{i=d+1}^m H^i_t \nonumber
\end{align}
where we denote $F_t := \psif_t \Bf_t, D^j_t:= \pdj_tD^j(t,T)$ and $H^i_t := \xi^i_t S^i_t$ and where we have used the repo trading constraint $\psi^i_t B^i_t + \xi^i_t S^i_t = 0$ for $i=1,2,\dots ,d$, which means that all long or short trades in the asset $S^i$ are funded using the account $B^i$ with the repo rate $\rilb$. Put another way, it is postulated that $H^i_t = \xi^i_t S^i_t = -  \psi^i_t B^i_t$ for $i=1,2,\dots ,d$.

The initial price $p_0$ received by the trader after he entered into an uncollateralized contract $A$ is equal both to his {\it initial wealth} $V_0(\phi ,A)$ and the {\it initial value} $V^p_0(\phi , A)$ of his portfolio. However, when a contract is collateralized, then the trader's initial wealth is still equal to $p_0$, but the initial value of the trader's portfolio equals $V^p_0(\phi , A) = p_0 +C_0$ where $C_0$ is the cash collateral, which is either pledged or received by the trader at time 0 (see Section \ref{sec3.1.2}). Our goal is to derive a probabilistic representation for the trader's unilateral initial price $p_0$ using replication-based arguments.

Assume that the $\gg$-adapted process $A$ of finite variation with $A_0=0$ represents the stream of cash flows (also known as the dividends) representing all future promised payoffs of a given contract. If a contract is non-defaultable and uncollateralized, then $A$ describes in fact all cash flows associated to the contract. Of course, the process $A$ needs to be complemented by additional cash flows when a contract is defaultable and collateralized. Let us stress that the initial price of the contract is not included in the cash flow stream $A$, since our goal is to derive the initial price from the future contract's cash flows, trading arrangements and model inputs. We first recall the definition of a self-financing trading strategy (also known as a dynamic portfolio) associated with a non-defaultable uncollateralized contract within the linear setup (see, for instance, Definition 2.3 in \cite{BR15}).

\bd \label{xdef1.1}
{\rm We say that a  trading strategy $\phi $ is {\it self-financing inclusive of a contract} $A$
if the value process $V^p(\phi,A )$ satisfies \eqref{sef0} and}
\begin{align}  \label{sef1}
dV^p_t (\phi,A ) &=  \psif_t \, d\Bf_t +\sum_{j=1}^2 \pdj_t\, dD^j(t,T) + \sum_{i=1}^d ( \psi^i_t \,  dB^i_t + \xi^i_t \,  dS^i_t )
 + \sum_{i=d+1}^m \xi^i_t\, dS^i_t + dA_t
 \\ & = \psif_t \, d\Bf_t +\sum_{j=1}^2 \pdj_t\, dD^j(t,T)  -  \sum_{i=1}^d \xi^i_t S^i_t (B^i_t)^{-1} \,  dB^i_t + \sum_{i=1}^d \xi^i_t \,  dS^i_t + \sum_{i=d+1}^m \xi^i_t\, dS^i_t + dA_t .  \nonumber
\end{align}
\ed

In view of the repo trading constraint $\xi^i_t S^i_t= -\psi^i_t B^i_t$ for $i=1,2,\dots , d$, we obtain
from \eqref{sef0}
\begin{equation}
dV^p_t (\phi,A ) = \rlb_t F_t \, dt +\sum_{j=1}^2 \pdj_t\, dD^j(t,T)  + \sum_{i=1}^d \xi^i_t ( dS^i_t - \rilb_t S^i_t\,  dt )
+ \sum_{l=d+1}^m \xi^i_t\, dS^i_t+ dA_t.
\end{equation}

\subsubsection{Non-defaultable Collateralized Contracts}   \lab{sec3.1.2}

Let us now consider the case of a collateralized version of a contract $A$, which is denoted as $(A,C)$ where the {\it margin account} $C$ is assumed to be an exogenously given $\gg$-semimartingale. We find it convenient to interpret the margin account as an additional stream of cash flows added to the process $A$, which was assumed to specify the `clean' (that is, uncollateralized and non-defaultable) version of the contract. For concreteness, we postulate that the cash collateral $C$ is rehypothecated, that is, it can be used for trading purposes. The process $C$ and its remuneration through accounts $B^{\pCc,l}$ and $B^{\pCc,l}$ (or, equivalently, the rates $\rcl$ and $\rcb$) are included in the definition of the process $A^C$ and thus they directly affect the dynamics of portfolio's value, as specified by the self-financing condition of Definition \ref{ahh4}. In the case of a non-defaultable collateralized contract with the margin process $C$ and collateral accrual rates $\rcl$ and $\rcb$ for the margin account, to compute the price and hedge for a collateralized contract, it suffices to replace the process $A$ by the process $A^C$ given by the following expression
\begin{align} \label{eqncc}
A^C_t &:= A_t + C_t + \int_0^t \pC_u^- (B^{\pCc,l}_u)^{-1}\, dB^{\pCc,l}_u -\int_0^t \pC_u^+ (B^{\pCc,b}_u)^{-1}\, dB^{\pCc,b}_u \nonumber \\ &= A_t + C_t + \int_0^t( \rcl_u \pC_u^- - \rcb_u \pC_u^+ ) \, du = A_t + C_t - \int_0^t \bar{c}_u C_u \, du
\end{align}
where we use the standard decomposition $C_t = C^+_t- C^-_t$ for every $t \in [0,T]$ and where we denote by $\bar{c}$ the effective collateral accrual rate, which equals
\be \lab{barc}
\bar{c}_t:= \rcl_t \I_{\{C_t<0\}} + \rcb_t \I_{\{C_t\geq 0\}}.
\ee
Hence the collateralized contract $(A,C)$ can be formally identified with the stream $A^C$ of cash flows.
Note that the process $V^p(\phi,A^C )$ is the {\it value process} of the trader's dynamic portfolio, whereas the process $V(\phi,A^C ) := V^p(\phi,A^C ) - C$ represents the trader's {\it wealth}. In particular, the terminal wealth satisfies $V_T(\phi,A^C ) = V^p_T(\phi,A^C )-C_T$.

We will now describe the dynamics of the value process of a self-financing trading strategy $\phi $ inclusive of cash flows
of a collateralized contract $(A,C)$. As was already mentioned, it suffices to extend Definition \ref{xdef1.1} to the case of
non-defaultable collateralized contracts.

\bd \lab{ahh4}
{\rm We say that a  trading strategy $\phi $ is {\it self-financing inclusive of a  non-defaultable collateralized contract} $(A,C)$
if the value process $V^p(\phi,A^C)$ satisfies \eqref{sef0} and}
\be \label{BSDE0}
dV^p_t (\phi,A^C ) = \rlb_t F_t \, dt+\sum_{j=1}^2 \pdj_t\, dD^j(t,T) + \sum_{i=1}^d \xi^i_t ( dS^i_t - \rilb_t S^i_t\,  dt )
 + \sum_{i=d+1}^m \xi^i_t\, dS^i_t + dA^C_t.
\ee
\ed

\subsubsection{Collateralized Defaultable Contracts} \lab{sec3.1.3}

In the next step, we introduce the concept of replication of a collateralized  defaultable contract up its effective maturity
date $\bartau = \tau \wedge T$. To this end, we first define the stream of cash flows for a  {\it collateralized defaultable contract} $(A,C,\recR,\tau )$. We set, for every $t \in [0,T]$,
\be \label{wtAC}
\wtAC_t=\I_{\{t < \tau \}} A_t+\I_{\{t \geq \tau \}} A_{\tau-} + \I_{\{t < \tau \}} C_t + \I_{\{t \geq \tau \}} C_{\tau-} - \int_0^{t \wedge \tau } \bar c_u C_u \, du
\ee
and
\be \label{wtACo}
\ACtheta_t=\wtAC_t + \I_{\{t \geq \tau \}} \recR_{\tau }.
\ee
The random variable $\recR_{\tau}$ represents a generic cash flow at the moment $\tau $ of the first default when it occurs either before or at $T$. We do not need to assume at this stage that $R_{\tau }$ is given by Definition \ref{close}. Then the process $\ACtheta$ gives all the cash flows of a defaultable collateralized contract $(A,C,\recR,\tau )$ up to its effective maturity $\bartau$.  We argue that Definition \ref{ahh4} can be extended in such a way that the process $\ACtheta$ gives the value process $V^p(\phi ,\ACtheta )$ on the stochastic interval $[0,\bartau ]$ and thus also the terminal value  $V^p_{\tau } (\phi , \ACtheta )$ on the event $\{\tau \leq T \}$.

\bd \lab{ahh5}
{\rm We say that a  trading strategy $\phi $ is {\it self-financing inclusive of a collateralized defaultable contract} $(A,C,\recR,\tau )$
if the value process $V^p(\phi, \ACtheta )$ satisfies on $[0,\bartau ]$ equality \eqref{sef0} and}
\be \label{BSDE0x}
dV^p_t (\phi,\ACtheta ) = \rlb_t F_t \, dt+\sum_{j=1}^2 \pdj_t\, dD^j(t,T) + \sum_{i=1}^d \xi^i_t ( dS^i_t - \rilb_t S^i_t\,  dt )
 + \sum_{i=d+1}^m \xi^i_t\, dS^i_t + d\ACtheta_t.
\ee
\ed

It is worth noting that $\ACtheta_t = \wtAC_t = A^C_t$ on the event $\{ t < \tau \}$ and thus the equalities
$$
V_t(\phi,\ACtheta )=V^p_t (\phi,\ACtheta)-C_t=V^p_t (\phi,\wtAC)-C_t=V^p_t(\phi,A^C)-C_t
$$
hold on the event $\{ t < \tau \}$ for every $t \in [0,T]$. Note that here $V^p(\phi,\wtAC)$ and $V^p(\phi,A^C)$ depend
on the cash flow $\recR_{\tau }$ through the initial price $p_0$ of the contract. These considerations lead to the following definition of the trader's wealth, in which we assume that the trader enters into a collateralized defaultable contract $(A,C,\recR,\tau )$ at the initial price $p_0$ and applies a self-financing strategy up to the contract's effective maturity $\bartau $.

\bd \label{wealth}
{\rm The {\it wealth process} $V(\phi, \ACtheta )$ of the trader equals, on the event $\{t< \tau \}$ for every $t \in [0,T]$,
\bde
V_t(\phi,\ACtheta )=V^p_t(\phi,\ACtheta )-C_t=V^p_t(\phi,\wtAC )-C_t
\ede
and on the event $\{ \tau \leq T \}$}
\bde
V_{\tau}(\phi,\ACtheta)=V^p_{\tau}(\phi,\ACtheta)=V^p_{\tau }(\phi,\wtAC )+\recR_{\tau }.
\ede
\ed

Note that Definition  \ref{wealth} is consistent with the fact that any particular specification the cash flow $\recR_{\tau }$
needs also to encompass the collateral either pledged or received be the trader just before the time of the first default, which is denoted as $C_{\tau-}$. We are now in a position to introduce the concept of a replicating strategy for a collateralized defaultable contract.

\bd \label{repl}
{\rm We say that a self-financing strategy $\phi $ {\it replicates} a collateralized defaultable contract $(A,C,\recR,\tau )$ if
$V_{\bartau }(\phi, \ACtheta ) = 0$ or, equivalently, the following equality holds
\be \label{wtACn}
V^p_{\bartau } (\phi, \wtAC ) = C_T \I_{\{ \tau > T\}} - \recR_{\tau } \I_{\{ \tau \leq T\}}.
\ee
Then the trader's {\it ex-dividend price} $\pi (A,C,\recR,\tau )$ for the collateralized defaultable contract $(A,C,\recR,\tau )$ is given by,
 on the event $\{ t < \tau \}$ for all $t \in [0,T]$,}
\be \label{exprice}
\pi_t(A,C,\recR,\tau ) = V_t(\phi,\ACtheta )=V^p_t(\phi, \wtAC) - C_t .
\ee
\ed

\brem \label{remrep}
It is clear that Definitions \ref{wealth} and \ref{repl} cover also the valuation of non-defaultable contracts. Formally, it suffices to postulate that $\tau >T$. In that case, we have that $V_t(\phi, A^C ) = V^p_t(\phi,A^C)-C_t$ for every $t \in [0,T]$ and replication of a contract $(A,C)$ means that $V_T(\phi , A^C)=0$ or, equivalently, that  $V^p_T(\phi , A^C)=C_T$.
\erem

\subsection{Valuation in a Linear Model} \lab{sec3.1.4}

For the sake of clarity of presentation, we first examine the valuation of non-defaultable  collateralized contracts and thus we now assume that defaultable bonds are not among traded assets. Also, to emphasize that the default times are not modeled, we denote the filtration by ${\mathbb F} = ({\cal F}_t)_{t \in [0,T]}$. In view of Definition \ref{repl} and Remark \ref{remrep},  the replication of a  collateralized contract $(A,C)$ by a self-financing trading strategy $\phi $ means that $V_T(\phi,A^C )= V^p_T(\phi,A^C )-C_T=0$.

\bd
{\rm The trader's {\it ex-dividend price} $\pi (A^C)$ for the collateralized contract $(A,C)$ is given by, for all $t \in [0,T]$,
$$
\pi_t(A,C) =  V_t(\phi,A^C ) = V^p_t (\phi,A^C )-C_t
$$
where $\phi $ is a self-financing trading strategy that replicates $(A,C)$.}
\ed

Assuming that a contract $(A,C)$ can be replicated, its initial trader's price $p_0 = \pi_0(A,C)$. In the next result, we have
that, for all $t \in [0,T]$,
\bde
V^p_t (\phi,A^C ) = F_t + \sum_{i=d+1}^m H^i_t
\ede
since we set $\kappa^1 = \kappa^2=0$ in equation \eqref{sef0}. For concreteness, we assume that $A_t = \I_{\{ t=T\}} X$ for all $t \in [0,T]$.

\bl \label{lemmcc}
Assume that the bonds $D^1$ and $D^2$ are not traded and $A_t = \I_{\{ t=T\}} X$ where $X$ is a square-integrable
$\F_T$-measurable random variable. Then the self-financing condition \eqref{BSDE0} yields the following dynamics for the trader's wealth process $V(\phi,A^C )$
 \begin{align} \label{eqnh7}
dV_t (\phi,A^C ) = \rlb_t \Big( V_t (\phi,A^C ) -  \sum_{i=d+1}^m H^i_t \Big)\, dt + \sum_{i=1}^d \xi^i_t ( dS^i_t -  h^i_t S^i_t\,  dt ) + \sum_{i=d+1}^m \xi^i_t\, dS^i_t + d\widehat C_t
\end{align}
where the process $\wh{C}$ is given by
\be \label{eqnh6a}
\wh{C}_t=\int_0^t (\rcl_u-\rlb_u)\pC_u^-\, du-\int_0^t(\rcb_u-\rlb_u)\pC_u^+\, du =\int_0^t (\rlb_u-\bar{c}_u )C_u\, du .
\ee
\el

\proof
Equality \eqref{eqnh7} follows from \eqref{eqncc},  \eqref{BSDE0} and the equality $V(\phi,A^C ) = V^p (\phi,A^C )-C$.
\endproof

From \eqref{eqnh7}, it is easy to derive the following linear BSDE for the trader's price process $\pi_t(X,C)=Y_t$
and the hedge ratios $\xi_t = Z_t$, for every $t \in [0,T)$,
\be  \label{BSDE1a}
dY_t = \Big( \rlb_t Y_t - \sum_{i=1}^d \rilb_t Z^i_t S^i_t - \sum_{i=d+1}^m \rlb_t Z^i_t S^i_t
+ (\rlb_t-\bar{c}_t )C_t \Big) \, dt + \sum_{i=1}^m Z^i_t\, dS^i_t
\ee
with the terminal condition $Y_T=-X$. Observe that the components $\psi^f, \psi^1, \dots , \psi^d$ of a self-financing trading strategy $\phi $, which replicates $(A,C)$, can also be computed from repo conditions $\xi^i_tS^i_t = -\psi^i_t B^i_t$ and equation \eqref{sef0}.

Under mild technical assumptions, the unique solution to the linear BSDE \eqref{BSDE1a} is known to exist in a suitable space
of stochastic processes and it is given by an explicit formula, of course, provided that the dynamics of $S^i$ are given.
For instance, if the prices of the risky assets $S^1,S^2,\dots ,S^m$ are governed by
$$
 dS^i_t = S^i_t \big( \mu^i_t \, dt + \sigma^i_t \, dW_t \big)
$$
where $W =(W^1,W^2,\dots,W^m)$ is an $m$-dimensional Brownian motion (possibly with the correlated components) with respect to its natural filtration ${\mathbb F}$, then \eqref{BSDE1a} becomes the classical linear BSDE (see, for instance, El Karoui et al. \cite{EPQ1997} or El Karoui and Quenez \cite{EQ1997})
\be  \label{BSDE1k}
dY_t = \Big( \rlb_t Y_t +\sum_{i=1}^d (\mu^i_t - \rilb_t ) Z^i_t S^i_t + \sum_{i=d+1}^m (\mu^i_t - \rlb_t ) Z^i_t S^i_t
+ (\rlb_t-\bar{c}_t )C_t \Big) \, dt + \sum_{i=1}^m Z^i_t \sigma^i_t \, dW_t
\ee
and thus an explicit expression for a solution $Y$ is known to exist under mild technical assumptions.  In Section \ref{sec3.1.6}, we  will derive, in particular, the following probabilistic representation for the ex-dividend price $\pi (X,C)$ in terms of the conditional expectation under a probability measure denoted as  $\Q^{f,h,f}$
\be \label{rnvavv}
\pi_t (X,C) = \Bf_t \, \mathbb{E}_{{\mathbb Q}^{f,h,f}} \bigg( - (\Bf_T)^{-1} X + \int_t^T  (\bar{c}_u - \rlb_u) C_u(\Bf_u)^{-1} \, du\, \Big| \, {\cal F}_t \bigg) .
\ee
Since $Y = \pi(X,C)$ is the solution to \eqref{BSDE1k}, the claimed representation \eqref{rnvavv} can also be obtained by an application of the Girsanov theorem and the well-known formula yielding an explicit solution to the linear BSDE. However, our general method will not rely on explicit solutions to BSDEs, but rather on abstract martingale arguments developed in Section  \ref{sec3.1.5}.

To illustrate the financial consequences of \eqref{rnvavv}, let us suppose, for instance, that $X \leq 0$ so that it is natural to assume that the collateral will always be pledged by the trader and thus $C_t = - C^-_t \leq 0$ for every $t \in [0,T]$. Then
\bde 
\pi_t(X,C)=\Bf_t\,\mathbb{E}_{{\mathbb Q}^{f,h,f}}\bigg(-(\Bf_T)^{-1}X-\int_t^T(c^l_u-\rlb_u)C^-_u(\Bf_u)^{-1}\,du\,\Big|\,{\cal F}_t\bigg)
\ede
and thus $\pi_t (X,C)  \geq \pi_t (X,0)$ provided that $c^l_t - \rlb_t \leq 0$ for every $t \in [0,T]$
where $\pi_t (X,0)$ is the trader's ex-dividend price of the uncollateralized version of the contract. This conclusion is consistent with the fact that it is unfavourable for the trader to borrow the cash amount $C^- \geq 0$ from the bank's treasury at the funding rate $f$ and to `lend' that amount to the counterparty in return for the remuneration at the lower rate $c^l \leq f$.

For concreteness, let us examine the valuation of a call option on the asset $S=S^1$. If the trader sells the call, then $X^1 = - (S_T-K)^+$ and thus the terminal condition for BSDE \eqref{BSDE1a} reads $Y_T = -X^1 =(S_T-K)^+$. This is, of course, consistent with the usual concept of replication of the payoff of the call by its writer. If the trader buys the call at time 0, then the terminal condition becomes $Y_T = -X^2 =-(S_T-K)^+ = -X^1$ since the terminal payoff from his perspective equals $X^2 = (S_T-K)^+$ and thus to hedge his exposure he needs to replicate the payoff $-(S_T-K)^+$. Note that the collateral $C^1_t =-(C^1_t)^-$ will be pledged at time $t$ by the trader who writes the call, but the collateral will be received if he buys the call, meaning that $C^2_t =(C^2_t)^+$ in the latter case.

Hence if $c^l \ne c^b$ then, even if we the postulate that $C^2_t =-C^1_t$ for all $t \in [0,T]$ (the natural symmetry of collateralization), we obtain the inequality
\bde
\pi_t (X^1,C^1) \ne  - \pi_t(-X^1,-C^1) = - \pi_t(X^2,C^2).
\ede
In contrast, the equalities
\bde
\pi_t (X^1,C^1) = - \pi_t(-X^1,-C^1) = - \pi_t(X^2,C^2)
\ede
are valid when that $c^l = c^b = c$ even if $c \ne f$. This illustrates the general property that the buying/selling trader's prices are equal in the linear setup provided that $c^l = c^b = c$ and under the postulate of symmetry of collateral. Formally, if $c^l = c^b = c$, then $\pi_t (A,C) = - \pi_t(-A,-C)$ for all $t \in [0,T]$ for an arbitrary specification of processes $A$ and $C$.

\subsubsection{Auxiliary Lemma} \lab{sec3.1.5}

To derive a general version of probabilistic representation \eqref{rnvavv}, we start by introducing the following notation
$$
B^{\zeta^j}_t := \exp \left( \int_0^t \zeta^j_u \, du \right), \ \ B^{\gamma^i }_t := \exp \left( \int_0^t \gamma^i_u \, du \right),\ \
B^{\nu^i }_t := \exp \left( \int_0^t \nu^i_u \, du \right)
$$
where $\zeta^j, \gamma^i $ and $\nu^i $ are arbitrary $\gg$-adapted and integrable processes. Then the processes
$\bar D^j(t,T) = (B^{\zeta^j }_t)^{-1} D^j(t,T),\, j=1,2$ satisfy
$$
d\bar D^j (t,T) = (B_t^{\zeta^j })^{-1} (dD^j(t,T) -  \zeta^j_t D^j(t,T) \, dt ).
$$
Then, for the risky assets traded under repo convention, we can define the processes $\bar S^i = (B^{\gamma^i })^{-1}S^i,\, i=1,2,\dots ,d $, so that we can write
$$
d\bar S^i_t = (B_t^{\gamma^i })^{-1} ( dS^i_t -  \gamma^i_t S^i_t \, dt ),\quad i=1,2,\dots ,d.
$$
Similarly, for the risky assets directly traded on the market, we can define $\bar S^i = (B^{\nu^i })^{-1}S^i ,\, i=d+1,d+2, \dots , m$, so that we get
$$
d\bar S^i_t = (B_t^{\nu^i })^{-1} ( dS^i_t -  \nu^i_t S^i_t \, dt ),\quad i=d+1,d+2, \dots , m.
$$

\bd \label{defmar}
{\rm Let $(\zeta , \gamma ,\nu ) = (\zeta^1, \zeta^2, \gamma^1, \gamma^2, \dots , \gamma^d, \nu^{d+1},\nu^{d+2},\dots , \nu^m)$ be an $(m+2)$-dimensional, $\gg$-adapted, integrable process. Then we denote by $\Q^{\zeta , \gamma ,\nu}$ a probability measure on $(\Omega ,{\cal G}_T)$ such that the processes $\bar D^j(t,T),\, j=1,2$ and $\bar S^i,\, i=1,2,\dots , m$ are $\Q^{\zeta,\gamma , \nu }$--local martingales.}
\ed

The existence of a probability measure $\Q^{\zeta,\gamma , \nu }$ is not obvious a priori, but it can be established in most market models encountered in the existing literature. For the sake of generality, we will henceforth postulate that such a probability measure is well defined. Then, from Definition \ref{defmar}, it follows that the processes
$$
D^j(t,T) -  \int_0^t \zeta^j_u D^j(u,T) \, du , \quad S^i_t - \int_0^t \gamma^i_u S^i_u \, du, \quad S^i_t - \int_0^t \nu^i_u S^i_u \, du
$$
are $(\Q^{\zeta ,\gamma ,\nu },\gg)$--local martingales. In other words, $\Q^{\zeta , \gamma ,\nu }$ is a local martingale measure for prices $D^j(t,T)$ discounted by  $B^{\zeta^j}$, the prices $S^i$ discounted with the processes $B^{\gamma^i }$ for $i=1,2, \dots , d$ and the prices $S^i$ discounted with the processes $B^{\nu^i }$ for $i=d+1,d+2, \dots , m$.

For an arbitrary $\gg$-adapted and integrable process $\eta$,  we define the process $B^\eta$ by setting, for every $t \in [0,T]$,
$$
B^{\eta }_t := \exp \left( \int_0^t \eta_u \, du \right).
$$
The following lemma underpins the probabilistic approach to the valuation of contracts under funding costs.

\bl \lab{lemb}
Assume that $V^p_t (\phi,\wtAC )$ is the value process of a self-financing trading strategy, in the sense of Definition \ref{ahh4},
so that \eqref{BSDE0} holds for $t \in [0,T]$ with $A^C = \wtAC$.
Let $\eta $ be an arbitrary $\gg$-adapted and integrable process and let the process $V^{\eta } (\phi,\wtAC ) $ be given by
\begin{align} \lab{gamm1y}
V^{\eta }_t (\phi,\wtAC ) :=\, & V^p_t (\phi,\wtAC ) + B^{\eta}_t \int_0^t \alpha_u F_u (B^{\eta }_u)^{-1} \, du
 + \sum_{j=1}^2 B^{\eta}_t \int_0^t \delta^j_u D^j_u (B^{\eta }_u)^{-1} \, du \\
& + \sum_{i=1}^d B^{\eta}_t \int_0^t  \beta^i_u H^i_u (B^{\eta }_u)^{-1} \, du
+ \sum_{i=d+1}^d B^{\eta}_t \int_0^t  \theta^i_u H^i_u (B^{\eta }_u)^{-1} \, du
 - B^{\eta}_t \int_{(0,t]} (B^{\eta }_u)^{-1} \, d\wtAC_u . \nonumber
\end{align}
If the following equalities hold for every $t \in [0,T]$
\be \lab{pronbcoeff}
\alpha_t=\eta_t - \rlb_t,\quad \delta^j_t=\eta_t-\zeta^j_t,\quad \beta^i_t=\eta_t -\gamma^i_t,\quad \theta^i_t=\eta_t-\nu^i_t,
\ee
then the process $\bar V^{\eta }(\phi,\wtAC ) := (B^{\eta })^{-1} V^{\eta }(\phi,\wtAC )$ is a $(\Q^{\zeta ,\gamma ,\nu },\gg)$-local martingale.
\el

\proof
Let us denote $V^{\eta}= V^{\eta }(\phi,\wtAC )$ and $V^p=V^p(\phi,\wtAC )$.
From equation \eqref{gamm1y}, we obtain
\bde
dV_t^{\eta} = dV^p_t + \alpha_t F_t \, dt  + \sum_{j=1}^2 \delta^j_t D^j_t \, dt  + \sum_{i=1}^d \beta^i_t H^i_t \, dt + ( V_t^{\eta} - V^p_t ) \eta_t \, dt - d\wtAC_t .
\ede
Since $V^p_t = F_t + \sum_{j=1}^2 D^j_t + \sum_{i=d+1}^m H^i_t$, we obtain
\begin{align*}
&dV_t^{\eta} - \eta_t V_t^{\eta} \, dt   =
 \rlb_t F_t \, dt  + \sum_{j=1}^2 \pdj_t \, dD^j(t,T) + \sum_{i=1}^d \xi^i_t ( dS^i_t - \rilb_t S^i_t\,  dt )+
  \sum_{i=d+1}^m \xi^i_t \, dS^i_t + \alpha_t F_t \, dt  \\
 & + \sum_{j=1}^2 \delta^j_t D^j_t \, dt+ \sum_{i=1}^d \beta^i_t H^i_t \, dt + \sum_{i=d+1}^m \theta^i_t H^i_t \, dt- \eta_t V^p_t \, dt \\
 & = (\alpha_t + \rlb_t - \eta_t ) F_t \, dt + \sum_{j=1}^2 (\delta^j_t + \zeta^j_t - \eta_t ) D^j_t \, dt +
 \sum_{j=1}^2 \pdj_t (dD^j(t,T) -  \zeta^j_t D^j(t,T) \, dt ) \\
 & + \sum_{i=1}^d (\beta^i_t + \gamma^i_t - \eta_t ) H^i_t \, dt + \sum_{i=1}^d \xi^i_t \, ( dS^i_t - \gamma^i_t S^i_t \, dt ) +  \sum_{i=d+1}^m (\theta^i_t + \nu^i_t - \eta_t ) H^i_t \, dt \\ & + \sum_{i=d+1}^m \xi^i_t \,( dS^i_t - \nu^i_t S^i_t \, dt ).
\end{align*}
Since $d\bar{V}^{\eta }_t = (B^{\eta}_t)^{-1}(dV_t^{\eta} - \eta_t V_t^{\eta} \, dt)$, it is now clear that if the processes $\alpha , \delta , \beta $ and $\theta $ satisfy \eqref{pronbcoeff}, then the process $\bar{V}^{\eta } =  (B^{\eta})^{-1}V^{\eta }$ is a
$(\Q^{\zeta ,\gamma ,\nu },\gg)$-local martingale.
\endproof

\subsubsection{Linear Probabilistic Valuation Formula} \lab{sec3.1.6}

We are in a position to prove the main result in this section, which gives the general probabilistic valuation formula.  We stress that the financial interpretations of processes $\eta , \zeta^1, \zeta^2, \gamma^i$ for $i=1,2, \dots ,d$  and $\nu^i$ for $i=d+1,d+2, \dots ,m$ are not relevant in the derivation of probabilistic representations \eqref{eqv1} and \eqref{eqv3} for the value process and the ex-dividend price, respectively. Let us denote, for all $t \in [0,T]$,
\be \label{eqnCA}
\wtA_t =\I_{\{t < \tau \}} A_t+\I_{\{t \geq \tau \}} A_{\tau-}, \quad \wtC_t =\I_{\{t < \tau \}} C_t+\I_{\{t \geq \tau \}} C_{\tau-}.
\ee
Recall also that the effective collateral accrual rate $\bar{c}$ for the margin account is given by \eqref{barc}.

\bt \label{rnv}
Assume that the collateralized defaultable contract $(A,C,\recR,\tau )$  can be replicated by a trading strategy $\phi $. If the associated process $\bar V^{\eta }(\phi,\wtAC )= (B^{\eta})^{-1}V^{\eta }(\phi , \wtAC )$, where $V^{\eta }(\phi , \wtAC )$ is given by \eqref{gamm1y}, is a true $(\Q^{\zeta ,\gamma ,\nu },\gg)$-martingale, then the value process $V^p_t(\phi,\ACtheta )$ equals,  on the event $\{ t < \tau \}$
for every $t \in [0,T]$,
 \begin{align} \label{eqv1}
&V^p_t (\phi,\ACtheta ) =B^{\eta}_t \, \mathbb{E}_{{\mathbb Q}^{\zeta, \gamma ,\nu }} \bigg(  - \int_{(t,{\bartau}]} (B^{\eta }_u)^{-1} \, d\wtAC_u + C_T (B^{\eta }_T)^{-1} \I_{\{ \tau > T \}} - \recR_{\tau } (B^{\eta }_{\tau } )^{-1} \I_{\{ \tau \leq T \}}
 \, \Big| \, {\cal G}_t \bigg) \nonumber
\\ &  + B^{\eta}_t \, \mathbb{E}_{{\mathbb Q}^{\zeta,\gamma ,\nu }} \bigg(
\int_t^{\bartau}  (\eta_u - \rlb_u) F_u (B^{\eta }_u)^{-1} \, du + \sum_{j=1}^2 \int_t^{\bartau} (\eta_u - \zeta^j_u) D^j_u (B^{\eta }_u)^{-1} \, du  \, \Big| \, {\cal G}_t \bigg)
\\ &  + B^{\eta}_t \, \mathbb{E}_{{\mathbb Q}^{\zeta,\gamma ,\nu }} \bigg(
\sum_{i=1}^d \int_t^{\bartau} (\rilb_u - \gamma^i_u) H^i_u (B^{\eta }_u)^{-1} \, du + \sum_{i=d+1}^m \int_t^{\bartau} (\eta_u - \nu^i_u) H^i_u (B^{\eta }_u)^{-1} \, du \, \Big| \, {\cal G}_t \bigg). \nonumber
\end{align}
Furthermore, the ex-dividend price for the collateralized defaultable contract $(A,C,\recR,\tau )$ is given by the following probabilistic expression, on the event $\{ t < \tau \}$ for every $t \in [0,T]$,
 \begin{align} \label{eqv3}
&\pi_t (A,C,\recR,\tau ) =B^{\eta}_t \, \mathbb{E}_{{\mathbb Q}^{\zeta, \gamma ,\nu }} \bigg( -\int_{(t,{\bartau}]} (B^{\eta }_u)^{-1} \, d\wtA_u
 - (\recR_{\tau } + C_{\tau-}) (B^{\eta }_{\tau } )^{-1} \I_{\{ \tau \leq T \}} \Big| \, {\cal G}_t \bigg)  \nonumber
  \\ &  + B^{\eta}_t \, \mathbb{E}_{{\mathbb Q}^{\zeta,\gamma ,\nu }} \bigg(\int_t^{\bartau}  (\eta_u - \rlb_u) F_u (B^{\eta }_u)^{-1} \, du\, + \sum_{j=1}^2 \int_t^{\bartau} (\eta_u - \zeta^j_u) D^j_u (B^{\eta }_u)^{-1} \, du  \, \Big| \, {\cal G}_t \bigg)
  \\ &  + B^{\eta}_t \, \mathbb{E}_{{\mathbb Q}^{\zeta,\gamma ,\nu }} \bigg(\sum_{i=1}^d \int_t^{\bartau} (\rilb_u - \gamma^i_u) H^i_u (B^{\eta }_u)^{-1} \, du+ \sum_{i=d+1}^m \int_t^{\bartau} (\eta_u - \nu^i_u) H^i_u (B^{\eta }_u)^{-1} \, du \, \Big| \, {\cal G}_t \bigg) \nonumber
   \\ &  + B^{\eta}_t \, \mathbb{E}_{{\mathbb Q}^{\zeta,\gamma ,\nu }} \bigg( \int_t^{\bartau}  (\bar{c}_u - \eta_u ) C_u  (B^{\eta }_u)^{-1}\, du\, \Big| \, {\cal G}_t \bigg) . \nonumber
\end{align}
\et

\proof
Let $\phi $ be a trading strategy which replicates the contract $(A,C,\recR,\tau )$. In view of Definition \ref{repl}, this means that
\be \label{termin}
V^p_{\bartau}(\phi ,\wt{A}^C) = C_T \I_{\{ \tau > T\}} - \recR_{\tau } \I_{\{ \tau \leq T\}}.
\ee
Equation \eqref{eqv1} is thus an immediate consequence of the martingale property of $\bar{V}^\eta (\phi,A^C )$, the fact that
$\bartau $ is a $\gg$-stopping time and equality \eqref{termin}. It remains to show that \eqref{eqv3} can be deduced from \eqref{eqv1} and the definition of the ex-dividend price.
On the event $\{t<\tau \}$, we have that $\pi_t(A,C,\recR,\tau )=V_t(\phi,\ACtheta )=V^p_t(\phi,\wtAC)-C_t $
(see \eqref{exprice}).  Note that \eqref{wtAC} gives $d\wtAC_t =d\wtA_t+d\wtC_t-\bar{c}_t C_t\,dt $.
The integration by parts formula yields, on the event $\{\tau > T\}$,
\begin{align*}
\int_{(t,\bartau ]} (B^{\eta }_u)^{-1} \, d\wtC_u &= \int_{(t,T]} (B^{\eta }_u)^{-1} \, dC_u= C_T (B^{\eta }_T)^{-1} - C_t (B^{\eta }_t)^{-1}
- \int_t^T C_u \, d(B^{\eta }_u)^{-1} \\
& =C_T (B^{\eta }_T)^{-1} - C_t (B^{\eta }_t)^{-1} + \int_t^T \eta_u (B^{\eta }_u)^{-1} C_u \, du
\end{align*}
whereas on the event $\{ \tau \leq T\}$, we obtain
\begin{align*}
\int_{(t,\bartau ]} (B^{\eta }_u)^{-1} \, d\wtC_u &= \int_{(t,\tau ]} (B^{\eta }_u)^{-1} \, d\wtC_u = C_{\tau-} (B^{\eta }_{\tau })^{-1} - C_t (B^{\eta }_t)^{-1} - \int_t^{\tau } C_u \, d(B^{\eta }_u)^{-1} \\
& = C_{\tau-} (B^{\eta }_{\tau })^{-1} - C_t (B^{\eta }_t)^{-1} + \int_t^{\tau } \eta_u (B^{\eta }_u)^{-1} C_u \, du.
\end{align*}
It is now easy to check that \eqref{eqv3} is a direct consequence of \eqref{eqv1}. \endproof

We will now consider some applications of Theorem \ref{rnv}. The first corollary gives an extension of equality \eqref{rnvavv} from
Section \ref{sec3.1.5}. It is arguably the most natural probabilistic representation for the ex-dividend price of a collateralized defaultable contract $(A,C,\recR,\tau )$. It should be noted that it gives a closed-form solution to the valuation problem and the right-hand side can be computed explicitly for several cases of interest.

\bcor \label{cor1.1}
If $\eta = f , \zeta^j = \rlb ,\, \gamma^i = h^i$ for $i=1,2,\dots , d$ and $\nu^i = \rlb $ for $i=d+1,d+2,\dots , m$, then \eqref{eqv1} yields the valuation formula inclusive of funding, defaults and liquidity costs
 \begin{align} \label{valff}
\pi_t (A,C,\recR,\tau ) &= \Bf_t \, \mathbb{E}_{{\mathbb Q}^{f,h,f}} \bigg( -\int_{(t,T]} (\Bf_u)^{-1} \, d\wtA_u
 - (\recR_{\tau } + C_{\tau-}) (\Bf_{\tau } )^{-1} \I_{\{ \tau \leq T \}} \, \Big| \, {\cal G}_t \bigg) \\
&+ \Bf_t \, \mathbb{E}_{{\mathbb Q}^{f,h,f}} \bigg(\int_t^{\bartau }  (\bar{c}_u - \rlb_u ) C_u  (\Bf_u)^{-1}\, du \, \Big| \, {\cal G}_t \bigg).  \nonumber
\end{align}
\ecor

Corollary \ref{cor1.1} is suitable when we are concerned with computation of the trader's price for $(A,C,\recR,\tau )$.
In contrast, it does not offer an immediate decomposition of the price $\pi_t (A,C,\recR,\tau )$ in terms of its `clean' price
and various {\it valuation adjustments}, which are of practical interest so that they were extensively studied in the
existing financial literature. For this reason, we will present in the next section another application of Theorem \ref{rnv}
in which the valuation adjustments, such as CVA, DVA, FVA, LVA, etc., will show explicitly.

\subsection{Risk-Neutral Valuation with Funding, Defaults and Liquidity Costs} \lab{sec3.2}

Our next goal is to show how to obtain from Theorem \ref{rnv} a version of the risk-neutral valuation formula with adjusted cash flows, which was first derived through different means in Pallavicini et al. \cite{PPB12} and Brigo et al. \cite{BFP16}. Let $r$ be a $\gg$-adapted, integrable process and let $B^r$ stand for the associated risk-free cash account
$$
B^r_t := \exp \left( \int_0^t r_u \, du \right).
$$
Let us stress that we do not assume that the risk-free cash account $B^r$ is available to the trader and thus the risk-free rate process $r$ can be seen by him as a purely instrumental variable.

\brem {\rm Although this is not required by our derivation, we can give to the rate $r$ the financial interpretation of a {\it risk-free rate}, that is, the short-term rate for lending/borrowing between non-defaultable entities. In practice, we can consider some market proxies.} In the European financial market, the OIS (Overnight Indexed Swap) rate can be seen as a proxy for the risk-free rate due to the fact that the counterparty risk is limited, as opposed to LIBOR, which is the rate of interest for cash lending/borrowing between defaultable entities. Therefore, the LIBOR-OIS spread is commonly interpreted as the indicator of an overall credit risk of LIBOR-based banks. In the United States, the corresponding interest spread refers to the LIBOR Eurodollar rate and the Federal Reserve's Fed Funds rate.
\erem

Upon setting $\eta = \zeta^j =  \gamma^i = \nu^i = r$ in \eqref{eqv3}, we obtain representation \eqref{eqv3x}, which is a version of the risk-neutral valuation with adjusted cash flows. To alleviate the notation, we henceforth write ${\mathbb Q}^{r}={\mathbb Q}^{r,r,r}$. Also, we write $F^i_t = \psi^i_t B^i_t = -H^i_t$

\bcor  \label{cor1.2}
The ex-dividend price for the collateralized defaultable contract $(A,C,\recR,\tau )$ is given by,  on the event $\{ t < \tau \}$
for every $t \in [0,T]$,
 \begin{align} \label{eqv3x}
&\pi_t (A,C,\recR,\tau ) =B^r_t \, \mathbb{E}_{{\mathbb Q}^{r}} \bigg( -\int_{(t,{\bartau}]} (B^r_u)^{-1} \, d\wtA_u
 - (\recR_{\tau } + C_{\tau-}) (B^r_{\tau } )^{-1} \I_{\{ \tau \leq T \}} \Big| \, {\cal G}_t \bigg)  \nonumber
  \\ &  + B^r_t \, \mathbb{E}_{{\mathbb Q}^{r}} \bigg(\int_t^{\bartau}  (r_u - \rlb_u) F_u (B^r_u)^{-1} \, du\,
   + \sum_{i=1}^d \int_t^{\bartau} (r_u - \rilb_u ) F^i_u (B^r_u)^{-1} \, du \, \Big| \, {\cal G}_t \bigg)
   \\ &  + B^r_t \, \mathbb{E}_{{\mathbb Q}^{r}} \bigg( \int_t^{\bartau}  (\bar{c}_u - r_u ) C_u  (B^r_u)^{-1}\, du\, \Big| \, {\cal G}_t \bigg) . \nonumber
\end{align}
\ecor

Observe that, by an application of Definition \ref{defmar}, the processes $\bar D^j(t,T) = (B^r_t)^{-1} D^j(t,T),\, j=1,2$
and $\bar S^i = (B_t^r)^{-1}S^i,\, i=1,2,\dots ,m$ or, equivalently, the processes
$$
D^j(t,T)-\int_0^t r_u D^j(u,T)\,du , \quad S^i_t-\int_0^t r_u S^i_u\,du,
$$
are $(\Q^{r},\gg)$--local martingales and thus the probability measure $\Q^r$ can be interpreted as a classical {\it risk-neutral probability}.

Let us consider some special cases of formula \eqref{eqv3x}. Suppose that the bank's treasury rate $\rlb $, the repo rates $h^i$, and the collateral accrual rates are all equal to the risk-free rate $r$. Then Corollary \ref{cor1.2} yields the following variant of the risk-neutral valuation formula for a collateralized defaultable contract
\be \label{rnvac1}
\pi_t (A,C,\recR,\tau ) = B^r_t \, \mathbb{E}_{{\mathbb Q}^{r}} \bigg( -\int_{(t,{\bartau}]}(B^r_u)^{-1} \, d\wtA_u  - (\recR_{\tau } + C_{\tau-}) (B^r_{\tau } )^{-1} \I_{\{ \tau \leq T \}}\, \Big| \, {\cal G}_t \bigg).
\ee
If, in addition, the equality  $\recR_{\tau } = - C_{\tau-}$ holds, meaning that the collateral is simply returned at the moment of the first default to the pledging party and no additional payoff occurs at time $\tau $ on the event $\{ \tau \leq T\}$, then \eqref{rnvac1} further reduces to
\begin{align} \label{rnvac2}
\pi_t (A,C,\recR,\tau ) & = B^r_t \, \mathbb{E}_{{\mathbb Q}^{r}} \bigg( -\int_{(t,{\bartau}]}(B^r_u)^{-1} \, d\wtA_u \, \Big| \, {\cal G}_t \bigg) \\
&= B^r_t \, \mathbb{E}_{{\mathbb Q}^{r}} \bigg( -  \I_{\{ \tau > T \}} \int_{(t,T]}(B^r_u)^{-1} \, dA_u
 -  \I_{\{ \tau \leq T \}} \int_{(t,\tau)}(B^r_u)^{-1} \, dA_u \, \Big| \, {\cal G}_t \bigg). \nonumber
\end{align}
Obviously, this elementary version of the risk-neutral valuation formula for collateralized defaultable contracts hinges on several assumptions, which are not satisfied in the currently prevailing market environment worldwide.

\subsubsection{CSA Closeout Valuation and the Payoff at the First Default}          \label{sec3.2.1}

Let us examine the consequences of the counterparty credit risk on the price of collateralized defaultable contract.
To this end, we need to specify the closeout valuation process $Q$ for the contract $(A,C,\recR,\tau )$. Let us stress that any specification for the process $Q$ should also include the promised payoff $A_{\tau } - A_{\tau-}$ at the moment of the first default. For instance, one could set (see \eqref{valff})
\bde
Q_t = B^f_t \, \mathbb{E}_{{\mathbb Q}^{f,h,f}} \bigg( -\int_{[t,T]}(B^f_u)^{-1} \, dA_u \, \Big| \, {\cal G}_t \bigg),
\ede
which would mean that the idiosyncratic funding costs of the trader would affect the CSA valuation.
A more conventional (and arguably more  appealing for practitioners,  although disputed by some researchers) form of the CSA closeout value hinges on the postulate that funding of all assets can be done at the risk-free rate $r$  and we henceforth follow this market convention. To be more specific, we set $f = h^i = r$ for $i=1,2,\dots ,d$ and $\tau > T$ in  \eqref{eqv3x} in order to obtain equality \eqref{defqq1}.  This means that the closeout valuation $Q$ is given by the `clean' price of the contract, that is, the price of the non-defaultable uncollateralized contract $A$ funded at the risk-free rate.

\bd \label{CSA}
{\rm The {\it risk-free closeout valuation} process $Q$ for the collateralized defaultable contract $(A,C,\recR,\tau )$ is given by
\be \label{defqq1}
Q_t := B^r_t \, \mathbb{E}_{{\mathbb Q}^{r}} \bigg( -\int_{[t,T]}(B^r_u)^{-1} \, dA_u \, \Big| \, {\cal G}_t \bigg)= \Delta A_t + \pi^r_t(A)
\ee
where $\Delta A_t = A_t - A_{t-}$ and $\pi_t^r(A),\, t \in [0,T]$ is the {\it ex-dividend risk-free price} of $A$, that is,}
\be \label{defqq2}
\pi^r_t(A) :=  B^r_t \, \mathbb{E}_{{\mathbb Q}^{r}} \bigg( -\int_{(t,T]}(B^r_u)^{-1} \, dA_u \, \Big| \, {\cal G}_t \bigg).
\ee
\ed

In view of Definition \ref{CSA}, at the moment of the first default we have, on the event $\{ \tau \leq T\}$,
\be  \label{defqq3}
Q_{\tau } =  A_{\tau } - A_{\tau -} +  B^r_{\tau} \, \mathbb{E}_{{\mathbb Q}^{r}} \bigg( -\int_{(\tau ,T]}(B^r_u)^{-1} \, dA_u \, \Big| \, {\cal G}_{\tau} \bigg) = \Delta A_{\tau } + \pi^r_{\tau }(A).
\ee

\subsubsection{Conventional Valuation Adjustments}          \label{sec3.2.2}

In this section, it is assumed that the recovery payoff $R_{\tau }$ is given by Definition \ref{close}.
For simplicity of presentation, we postulate, in addition, that the event $\{\tau_C = \tau_I\}$ is negligible under $\Q^r$ and thus the closeout payoff $\theta_{\tau }$ is given by \eqref{deftt} with $Q$ given by \eqref{defqq1}. Then we have following result, which is a consequence of Corollary \ref{cor1.2} and Definition \ref{CSA}.

\bp  \label{pro1.2}
 Under the risk-free closeout valuation $\epsiQ$ given by \eqref{defqq1}, the ex-dividend price for the collateralized defaultable contract $(A,C,\recR,\tau )$ equals,  on the event $\{ t < \tau \}$ for every $t \in [0,T]$,
 \begin{align} \label{eqv3m}
&\pi_t (A,C,\recR,\tau ) = \pi^r_t (A) + B^r_t \, \mathbb{E}_{{\mathbb Q}^{r}} \Big(
\I_{\{ \tau \leq T\}} \big(  \I_{\{\tau_C< \tau_I \}} L_C \Upsilon^+ - \I_{\{\tau_I < \tau_C \}} L_I \Upsilon^- \big) \, \Big| \, {\cal G}_t \Big)  \nonumber
  \\ &  + B^r_t \, \mathbb{E}_{{\mathbb Q}^{r}} \bigg(\int_t^{\bartau}  (r_u - \rlb_u) F_u (B^r_u)^{-1} \, du
  + \sum_{i=1}^d \int_t^{\bartau} (r_u -\rilb_u ) F^i_u (B^r_u)^{-1} \, du
   \, \Big| \, {\cal G}_t \bigg)
   \\ &  + B^r_t \, \mathbb{E}_{{\mathbb Q}^{r}} \bigg( \int_t^{\bartau}  (\bar{c}_u - r_u ) C_u  (B^r_u)^{-1}\, du\, \Big| \, {\cal G}_t \bigg) . \nonumber
\end{align}
\ep

\proof
In view of Definition \ref{CSA}, to obtain \eqref{eqv3m} from \eqref{eqv3x} and \eqref{deftt}, it suffices to
observe that, on the event $\{t < \tau \}$,
\bde
\int_{(t,{\bartau}]} (B^r_u)^{-1} \, d\wtA_u  = \int_{(t,T]} (B^r_u)^{-1} \, dA_u - \I_{\{ \tau \leq T\}}  \int_{[\tau,T]} (B^r_u)^{-1} \, dA_u
\ede
and thus
\begin{align*}
 B^r_t \, \mathbb{E}_{{\mathbb Q}^{r}} \bigg( - \int_{(t,{\bartau}]} (B^r_u)^{-1} \, d\wtA_u \, \Big| \, {\cal G}_t \Big)  &
 = B^r_t \, \mathbb{E}_{{\mathbb Q}^{r}} \bigg( - \int_{(t,T]} (B^r_u)^{-1} \, dA_u + \I_{\{ \tau \leq T\}}  \int_{[\tau,T]} (B^r_u)^{-1} \, dA_u  \, \Big| \, {\cal G}_t \Big) \\ &= \pi^r_t (A)  + \I_{\{ \tau \leq T\}} Q_{\tau }
\end{align*}
where the last equality follows from \eqref{defqq1} and \eqref{defqq2}.
\endproof

Representation \eqref{eqv3m} of the trader's ex-dividend price for $(A,C,\recR,\tau )$ can be given the following financial interpretation
\begin{align} \label{xvaformula}
\pi_t (A,C,\recR,\tau )=\pi^r_t (A)+\CVA_t-\DVA_t+\FVA^f_t+\sum_{i=1}^m \FVA^{h^i}_t+\LVA_t
\end{align}
where the {\it credit valuation adjustment} $\CVA_t$ equals
\bde 
\CVA_t=B^r_t\,\mathbb{E}_{{\mathbb Q}^{r}}\Big(\I_{\{\tau \leq T\}}\I_{\{\tau_C<\tau_I\}}L_C\Upsilon^+\,\Big|\,{\cal G}_t\Big),
\ede
the {\it debit valuation adjustment} $\DVA_t$ equals
\bde 
\DVA_t=B^r_t\,\mathbb{E}_{{\mathbb Q}^{r}}\Big(\I_{\{ \tau \leq T\}}\I_{\{\tau_I < \tau_C \}}L_I\Upsilon^-\,\Big|\,{\cal G}_t\Big),
\ede
the {\it treasury funding valuation adjustment} $\FVA^f_t$ is given by
\bde 
\FVA^f_t=B^r_t\,\mathbb{E}_{{\mathbb Q}^{r}}\bigg(\int_t^{\bartau}(r_u - \rlb_u)F_u (B^r_u)^{-1}\,du\,\Big| \, {\cal G}_t \bigg),
\ede
the {\it repo funding valuation adjustments} $\FVA^{h^i}_t$ for $i=1,2, \dots , d$ are given by
\bde 
\FVA^{h^i}_t = B^r_t\,\mathbb{E}_{{\mathbb Q}^{r}}\bigg(\int_t^{\bartau}(r_u-\rilb_u) F^i_u(B^r_u)^{-1}\, du\,\Big|\,{\cal G}_t \bigg),
\ede
and  {\it liquidity valuation adjustment} $\LVA_t$ satisfies
\bde 
\LVA_t= B^r_t\,\mathbb{E}_{{\mathbb Q}^{r}}\bigg(\int_t^{\bartau}(\bar{c}_u - r_u )C_u(B^r_u)^{-1}\, du\, \Big| \, {\cal G}_t \bigg).
\ede

\ssc{Nonlinear Models with Funding Costs and Defaults}  \lab{sec3.3}

A nonlinear extension of the linear framework introduced in the preceding section is obtained when a single cash rate $\rlb$ is replaced by differential lending and borrowing rates, denoted as $\rll$ and $\rbb$, respectively, and, similarly, by introducing differential repo rates for long and short positions in the stock $S^i$, which are denoted as $\rill$ and $\ribb$, respectively.
It is worth mentioning that, for the sake of conciseness, we do not deal here with external funding adjustments introduced in Section  \ref{sec2.7}. Hence Proposition \ref{pro1.2b}, as well as the price decomposition given by \eqref{xvaformulan}, should be seen as  extensions of Lemma \ref{lemm2.1}, rather than Proposition \ref{profin}. Then we have the following representation of the value process $V^p(\phi,\ACtheta )$ of a portfolio $\phi$
\bde
V^p_t (\phi, \ACtheta ) =  \psi^l_t B^l_t + \psi^b_t B^{i,b}_t + \sum_{j=1}^2 \pdj_tD^j(t,T) +
\sum_{i=1}^d \big( \psi^{i,l}_t B^{i,l}_t + \psi^{i,b}_t B^{i,b}_t + \xi^i_t S^i_t \big) + \sum_{i=d+1}^m \xi^i_t  S^i_t .
\ede
Consistently with the financial interpretation of trading within the nonlinear framework, we postulate that $\psi^l_t \geq 0, \psi^b_t \leq 0$ and $\psi^l_t \psi^b_t = 0$ for all $t \in [0,T]$. Similarly, we assume that $\psi^{i,l}_t \geq 0,  \psi^{i,b}_t \leq 0$ and $\psi^{i,l}_t \psi^{i,b}_t = 0$ for $i=1,2,\dots ,d$ and, in addition, we impose the repo trading condition for risky assets $S^1,S^2,\dots ,S^d$
\bde
\psi^{i,l}_t B^{i,l}_t + \psi^{i,b}_t B^{i,b}_t + \xi^i_t S^i_t = 0 .
\ede
For convenience, we denote
\begin{align*}
F_t &= \psi^l_t B^l_t + \psi^b_t B^b_t = V^p_t (\phi, \ACtheta ) - \sum_{j=1}^2 \pdj_tD^j(t,T) - \sum_{i=d+1}^m \xi^i_t S^i_t \\
  &= V^p_t (\phi, \ACtheta ) - \sum_{j=1}^2 D^j_t - \sum_{i=d+1}^m H^i_t
\end{align*}
where $D^j_t =\pdj_tD^j(t,T)$ for $j=1,2$ and $H^i_t = \xi^i_t S^i_t = - F^i_t$ for $i=1,2,\dots, m$.

\sssc{Nonlinear Dynamics of the Value Process of a Trading Strategy} \label{sec3.3.1}

We are now in a position to derive the non-linear dynamics of the value process and thus also obtain in Section \ref{sec3.3.4}
a generic non-linear pricing BSDE for a collateralized defaultable contract $(A,C,\recR,\tau )$.
Recall that $\tau = \tau_I \wedge \tau_C $ so that $\bartau := \tau \wedge T$  is the effective maturity of the contract.

\bl \label{lemcc}
We have $\psi^{l}_t = (B^{l}_t)^{-1} F_t^+ , \, \psi^{b}_t = -(B^{b}_t)^{-1} F_t^-$
and for $i=1,2,\dots , d$
\bde
\psi^{i,l}_t =  (B^{i,l}_t)^{-1} (H^i_t)^-=  (B^{i,l}_t)^{-1} (F^i_t)^+, \quad 
\psi^{i,b}_t = -(B^{i,b}_t)^{-1} (H^i_t)^+ = -(B^{i,b}_t)^{-1} (F^i_t)^-. 
\ede
\el

\proof
It suffices to note that
\bde
\psi^l_t B^l_t + \psi^b_t B^b_t = F_t , \quad
\psi^{i,l}_t B^{i,l}_t + \psi^{i,b}_t B^{i,b}_t = - \xi^i_t S^i_t = - H^i_t = F^i_t
\ede
and to make use of the postulated conditions.
\endproof


\bl \label{lemcxz}
The value process $V^p(\phi,\ACtheta )$ of a self-financing trading strategy associated with
collateralized defaultable contract $(A,C,\recR,\tau )$ satisfies on $[0,\bartau ]$
\begin{align} \lab{xzx}
dV^p_t (\phi,\ACtheta ) =\, & \rll_t F_t^+ \, dt - \rbb_t F_t^- \, dt +\sum_{j=1}^2 \pdj_t \,dD^j(t,T) + \sum_{i=1}^d \Big( \xi^i_t \, dS^i_t + \rill_t (F^i_t)^+ \, dt - \ribb_t (F^i_t)^- \,  dt \Big) \nonumber  \\ &
 + \sum_{i=d+1}^m \xi^i_t\, dS^i_t
   + d\wtA_t + d\wtC_t - \bar{c}_tC_t\, dt  + d \big( \I_{\{ t \geq \tau \}} \recR_{\tau } \big)
\end{align}
where $\bar{c}$ is given by \eqref{barc},  $\wtA$ and $\wtC$ are given by \eqref{eqnCA} and $\recR_{\tau }$ is given by \eqref{closeout}.
\el

\proof
By a slight extension of Definition \ref{ahh5}, the self-financing condition for $\phi$ reads, for $t \in [0,\bartau ]$,
\begin{align*}
dV^p_t (\phi,\ACtheta ) =\,&  \psi^l_t \, dB^l_t + \psi^b_t\, dB^b_t + \sum_{j=1}^2 \pdj_t\,dD^j(t,T)+
\sum_{i=1}^d ( \psi^{i,l}_t \, dB^{i,l}_t + \psi^{i,b}_t \,  dB^{i,b}_t + \xi^i_t \, dS^i_t )
\\ &  + \sum_{i=d+1}^m \xi^i_t \, dS^i_t + d\ACtheta_t
\end{align*}
where $\ACtheta$ is given by \eqref{wtACo}. Therefore, in view of Lemma  \ref{lemcc}, we obtain the following expression for the dynamics of the value process
\begin{align*}
dV^p_t (\phi,\ACtheta ) =  \, &(B^{l}_t)^{-1} F_t^+ \, dB^l_t -(B^{b}_t)^{-1} F_t^- \, dB^b_t
 +\sum_{j=1}^2 \pdj_t\,dD^j(t,T)+ \sum_{i=d+1}^m \xi^i_t \, dS^i_t \\
 & + \sum_{i=1}^d (B^{i,l}_t)^{-1} (F^i_t)^+ \, dB^{i,l}_t- \sum_{i=1}^d (B^{i,b}_t)^{-1} (F^i_t)^- \,  dB^{i,b}_t + \sum_{i=1}^d \xi^i_t \, dS^i_t +  d\ACtheta_t .
\end{align*}
Under the assumptions of an absolute continuity of cash/repo accounts, using also \eqref{wtACo} and \eqref{eqnCA},  we obtain \eqref{xzx}.
\endproof

To obtain a  convenient linearized representation of the dynamics of $V^p(\phi,\wtAC )$, we introduce the {\it effective funding rate} $\bar \rlb$ and the {\it effective repo rates} $\bar{h}^i$. It is obvious that effective rate depend on the trader's strategy $\phi$, in general.

\bl \label{lemmbc}
The process $V^p(\phi,\wtAC )$ satisfies on $[0,\bartau ]$
\begin{align} \lab{tgb}
dV^p_t (\phi,\wtAC ) =( \bar \rlb_t F_t + \bar{h}^i_t F^i_t - \bar{c}_tC_t ) \, dt +\sum_{j=1}^2 \pdj_t \, dD^j(t,T) + \sum_{i=1}^m \xi^i_t \, dS^i_t  + d\wtA_t + d\wtC_t
\end{align}
where the effective funding rate $\bar \rlb$ equals
\be \lab{rr1}
\bar \rlb_t := \rll_t \,\I_{\{ F_t \geq 0\}}  + \rbb_t \, \I_{\{ F_t < 0\}}
\ee
and the effective repo rates $\bar{h}^i,\, i=1,2,\dots ,d$ are given by
\be \lab{rr2}
\bar{h}^i_t := \rill_t \, \I_{\{F^i_t \geq 0\}}  + \ribb_t \, \I_{\{ F^i_t < 0\}} .
\ee
\el

\sssc{Nonlinear Probabilistic Valuation Formula} \label{sec3.3.2}

Recall that
$$
B^{\eta}_t := \exp \left( \int_0^t \eta_u \, du \right), \quad B^{\zeta^j}_t := \exp \left( \int_0^t \zeta^j_u \, du \right), \quad
B^{\gamma^i}_t := \exp \left( \int_0^t \gamma^i_u \, du \right)
$$
where  $\eta , \zeta^1, \zeta^2 , \gamma^i,\, i=1,2, \dots ,d$ and $\nu^i ,\, i=d+1,d+2, \dots , m$ are arbitrary adapted and integrable processes. Let ${\mathbb Q}^{\zeta ,\gamma , \nu}$ be a probability measure such that the processes $(B^{\zeta^j })^{-1}D^j(t,T),\, j=1,2,\, (B^{\gamma^i })^{-1}S^i,\, i=1,2,\dots ,d$ and $(B^{\nu^i })^{-1}S^i,\, i=d+1,d+2,\dots ,m$  are ${\mathbb Q}^{\zeta , \gamma ,\nu }$-local martingales.

Then Lemmas \ref{lemb} and \ref{lemmbc} yield the following result, which is a nonlinear counterpart of Theorem \ref{rnv}.
For the validity of this result, one needs to impose some mild integrability assumptions.
Recall from Definition \ref{close} that the CSA closeout payoff satisfies $\theta_{\tau } = \recR_{\tau } + C_{\tau -}.$

\newpage

\bt \label{rnve}
Assume that the collateralized defaultable contract $(A,C,\recR,\tau )$  can be replicated by a trading strategy $\phi $ and the associated process $\bar V^{\eta }(\phi,\wtAC )$ is a true martingale under $\Q^{\zeta ,\gamma ,\nu }$. Then $V^p_t(\phi,\ACtheta )$ equals, on the event $\{ t < \tau \}$
 for every $t \in [0,T]$,
 \begin{align} \label{eqv1nl}
&V^p_t (\phi,\ACtheta ) =B^{\eta}_t \, \mathbb{E}_{{\mathbb Q}^{\zeta, \gamma ,\nu }} \bigg( - \int_{(t,{\bartau}]} (B^{\eta }_u)^{-1} \, d\wtAC_u + C_T (B^{\eta }_T)^{-1} \I_{\{ \tau > T \}} - \recR_{\tau } (B^{\eta }_{\tau } )^{-1} \I_{\{ \tau \leq T \}}
 \, \Big| \, {\cal G}_t \bigg) \nonumber
\\ &  + B^{\eta}_t \, \mathbb{E}_{{\mathbb Q}^{\zeta,\gamma ,\nu }} \bigg(
\int_t^{\bartau}  (\eta_u - \bar{\rlb}_u) F_u (B^{\eta }_u)^{-1} \, du + \sum_{j=1}^2 \int_t^{\bartau} (\eta_u - \zeta^j_u) D^j_u (B^{\eta }_u)^{-1} \, du  \, \Big| \, {\cal G}_t \bigg)
\\ &  + B^{\eta}_t \, \mathbb{E}_{{\mathbb Q}^{\zeta,\gamma ,\nu }} \bigg(
\sum_{i=1}^d \int_t^{\bartau} ( \bar{h}^i_u - \gamma^i_u) H^i_u (B^{\eta }_u)^{-1} \, du + \sum_{i=d+1}^m \int_t^{\bartau} (
\eta_u - \nu^i_u) H^i_u (B^{\eta }_u)^{-1} \, du \, \Big| \, {\cal G}_t \bigg). \nonumber
\end{align}
The ex-dividend price for the collateralized defaultable contract $(A,C,\recR,\tau )$ is given by,  on the event $\{ t < \tau \}$ for every $t \in [0,T]$,
 \begin{align} \label{eqv3nl}
&\pi_t (A,C,\recR,\tau ) =B^{\eta}_t \, \mathbb{E}_{{\mathbb Q}^{\zeta, \gamma ,\nu }} \bigg( -\int_{(t,{\bartau}]} (B^{\eta }_u)^{-1} \, d\wtA_u
 - \theta_{\tau } (B^{\eta }_{\tau } )^{-1} \I_{\{ \tau \leq T \}}  + \int_t^{\bartau}  (\eta_u - \bar{\rlb}_u) F_u (B^{\eta }_u)^{-1} \, du\, \Big| \, {\cal G}_t \bigg)  \nonumber
  \\ &  + B^{\eta}_t \, \mathbb{E}_{{\mathbb Q}^{\zeta,\gamma ,\nu }} \bigg(\sum_{j=1}^2 \int_t^{\bartau} ( \eta_u - \zeta^j_u) D^j_u (B^{\eta }_u)^{-1} \, du + \sum_{i=1}^d \int_t^{\bartau} ( \bar{h}^i_u - \gamma^i_u) H^i_u (B^{\eta }_u)^{-1} \, du \, \Big| \, {\cal G}_t \bigg)
  \\ &  + B^{\eta}_t \, \mathbb{E}_{{\mathbb Q}^{\zeta,\gamma ,\nu }} \bigg( \sum_{i=d+1}^m \int_t^{\bartau} ( \eta_u - \nu^i_u) H^i_u (B^{\eta }_u)^{-1} \, du + \int_t^{\bartau}  (\bar{c}_u - \eta_u ) C_u  (B^{\eta }_u)^{-1}\, du\, \Big| \, {\cal G}_t \bigg) \nonumber
\end{align}
where the effective rates $\bar{\rlb}, \bar{h}^i$ and $\bar{c}$ are given by \eqref{rr1}, \eqref{rr2} and \eqref{barc}, respectively.
\et

\proof
In view of Lemma \ref{lemmbc}, the proof of Theorem \ref{rnve} is analogous to the proof of Theorem \ref{rnv} and thus it is omitted.
\endproof

\sssc{Nonlinear Risk-Neutral Valuation with Funding, Defaults and Liquidity} \lab{sec3.3.3}

Theorem \ref{rnve} can be used to identify contributions of various adjustments to the trader's price with respect to the {\it clean price} computed using the short-term rate. To this end, one can use the following consequence of  Theorem \ref{rnve}, which corresponds to Corollary \ref{cor1.2}. Recall that we use the shorthand notation ${\mathbb Q}^{r}={\mathbb Q}^{r,r,r}$. As in Section \ref{sec3.2.2}, we assume that the event $\{\tau_C = \tau_I\}$ is negligible under ${\mathbb Q}^{r}$  and thus, without loss of generality, the CSA closeout payoff $\theta_{\tau }$ is assumed to be given by \eqref{deftt}.

\bcor \label{noncor}
The ex-dividend price for the collateralized defaultable contract $(A,C,\recR,\tau )$ satisfies, on the event $\{ t < \tau \}$ for every $t \in [0,T]$,
 \begin{align} \label{eqtv3x}
&\pi_t (A,C,\recR,\tau ) = B^r_t \, \mathbb{E}_{{\mathbb Q}^{r}} \bigg( -\int_{(t,{\bartau}]} (B^r_u)^{-1} \, d\wtA_u
 - \theta_{\tau } (B^{\eta }_{\tau } )^{-1} \I_{\{ \tau \leq T \}} + \int_t^{\bartau}  (r_u - \bar{\rlb}_u) F_u (B^r_u)^{-1} \, du\, \Big| \, {\cal G}_t \bigg)   \nonumber
  \\ &  + B^r_t \, \mathbb{E}_{{\mathbb Q}^{r}} \bigg( \sum_{i=1}^d \int_t^{\bartau} ( r_u - \bar{h}^i_u ) F^i_u (B^r_u)^{-1} \, du  + \int_t^{\bartau}  (\bar{c}_u - r_u ) C_u  (B^r_u)^{-1}\, du\, \Big| \, {\cal G}_t \bigg) .
\end{align}
\ecor

We maintain the assumption made in Section \ref{sec3.2.1} about the exact specification of the CSA closeout payoff $\theta $.
Recall also that the case of the joint defaults was precluded.  Then Proposition \ref{pro1.2} can be easily extended to cover the
case of the nonlinear setup. It suffices to replace $f,h^i$ and $c$ by $\bar{f}, \bar{h}^i$ and $\bar{c}$, respectively, in equation \eqref{eqv3m}. The proof of the next result is analogous to the proof of  Proposition \ref{pro1.2}  and thus it is omitted.

\newpage

\bp  \label{pro1.2b}
Under the risk-free closeout valuation $\epsiQ$ given by \eqref{defqq1}, the ex-dividend price for the collateralized defaultable contract $(A,C,\recR,\tau )$ equals, on the event $\{ t < \tau \}$ for every $t \in [0,T]$,
 \begin{align} \label{eqv3mc}
&\pi_t(A,C,\recR,\tau ) = \pi^r_t(A) + B^r_t \, \mathbb{E}_{{\mathbb Q}^{r}} \Big( \I_{\{ \tau \leq T\}} \big(  \I_{\{\tau_C< \tau_I \}} L_C \Upsilon^+ - \I_{\{\tau_I < \tau_C \}} L_I \Upsilon^- \big) \, \Big| \, {\cal G}_t \Big)  \nonumber \\ &
+ B^r_t \, \mathbb{E}_{{\mathbb Q}^{r}} \bigg(\int_t^{\bartau}  (r_u - \bar{\rlb}_u) F_u (B^r_u)^{-1} \, du + \sum_{i=1}^d \int_t^{\bartau}
(r_u-\bar{h}^i_u) F^i_u (B^r_u)^{-1} \, du \, \Big| \, {\cal G}_t \bigg) \\ & + B^r_t \, \mathbb{E}_{{\mathbb Q}^{r}} \bigg( \int_t^{\bartau}  (\bar{c}_u - r_u ) C_u  (B^r_u)^{-1}\, du\, \Big| \, {\cal G}_t \bigg) .\nonumber
\end{align}
\ep

Proposition \ref{pro1.2b} leads in turn to the following formal decomposition of the ex-dividend price for $(A,C,\recR,\tau )$ into its `clean' price and valuation adjustments (for their interpretation, see Section \ref{sec3.2.2}), which holds on the event $\{ t < \tau \}$ for every $t \in [0,T]$,
\begin{align} \label{xvaformulan} 
\pi_t(A,C,\recR,\tau )=\pi^r_t(A)+\CVA_t-\DVA_t+\FVA^{\bar{f}}_t+\sum_{i=1}^m \FVA^{\bar{h}^i}_t+\LVA_t .
\end{align}

\sssc{Nonlinear BSDE for Valuation and Hedging} \lab{sec3.3.4}

As opposed to the linear setup, Theorem \ref{rnve} and Corollary \ref{noncor} do not furnish closed-form expressions for the trader's price, since their right-hand sides involve several unknown processes. In fact, the associated BSDE is nonlinear so its solution (if it exists and is unique) is not known explicitly. Under the standing assumption that the trader's initial endowment is null, equation \eqref{tgb} leads to the following nonlinear BSDE for the portfolio's value $Y := V^p(\phi,\wtAC )$ and hedge ratios $(U,Z) := (\kappa , \xi )$ (the remaining components of the replicating strategy $\phi $ can be found using Lemma \ref{lemcc}). Then that the equality $\pi_t (A,C,\recR,\tau )=Y_t -C_t$ holds on the event $\{ t < \tau \}$ for every $ t\in [0,T]$.

\bp
The portfolio's value $Y = V^p_t (\phi,\wtAC )$ and the hedge ratios $(U,Z) = (\kappa , \xi )$ satisfy
\begin{align*} 
dY_t = \Big( \bar \rlb_t \wh{Y}_t -  \sum_{i=1}^d \bar{h}^i_t Z^i_t S^i_t - \bar{c}_tC_t \Big)\, dt +\sum_{j=1}^2 U^j_t\,dD^j(t,T) + \sum_{i=1}^m Z^i_t \, dS^i_t + d\wtA_t + d\wtC_t
\end{align*}
where
\bde
\wh{Y}_t := Y_t - \sum_{j=1}^2 U^j_t D^j(t,T) - \sum_{i=d+1}^m Z^i_t S^i_t
\ede
and  from \eqref{rr1}--\eqref{rr2}
\begin{align*} 
&\bar \rlb_t := \rll_t \,\I_{\{ \wh{Y}_t \geq 0\}}  + \rbb_t \, \I_{\{ \wh{Y}_t < 0\}}, \\
&\bar{h}^i_t := \rill_t \, \I_{\{ Z^i_t S^i_t < 0\}}  + \ribb_t \, \I_{\{ Z^i_t S^i_t \geq 0\}}, \\
&\bar{c}_t:= \rcl_t \I_{\{C_t<0\}} + \rcb_t \I_{\{C_t\geq 0\}}.
\end{align*}
The terminal condition at $\bartau $  reads  $Y_{\bartau} = \I_{\{ \tau > T \}} C_T  - \I_{\{ \tau \leq T \}} R_{\tau }$
where the recovery payoff $\recR_{\tau }$ equals $($see \eqref{wtACn} and \eqref{closeout}, respectively$)$
\bde
\recR_{\tau } = \I_{\{\tau_C< \tau_I \}}(R_C \Upsilon^+-\Upsilon^-) + \I_{\{\tau_I < \tau_C \}}(\Upsilon^+ - R_I \Upsilon^-)
\ede
where $\Upsilon = Q_{\tau} - C_{\tau-}$.
\ep

Given any specific semimartingale model for risky assets, it is usually possible to show that the pricing BSDE has
a unique solution in a suitable space of stochastic processes. For instance, Nie and Rutkowski \cite{NR1,NR2,NR4} examine
the valuation and hedging of a contract $(A,C)$ (so defaults are not considered) with both an exogenous and an endogenous collateralization.


\sect{Market Incompleteness}\label{sec4}

In the previous section, we have assumed that the contract that we are evaluating can be replicated. Now we wish to examine briefly what happens if this assumption is relaxed. We will concentrate here on the case where the contract is subject to defaults of both the bank and the counterparty, while the bank cannot trade its own or counterparty's bonds (i.e., $\kappa^1=\kappa^2=0$). To further simplify the setup, we suppose that all of the hedging instruments are traded on the repo market (i.e., $d=m$) and that repo rates are the same across all positions and directions of the trade (i.e., $h^{i,l}=h^{i,b}=h$ for all $i$). Furthermore, we will aim to reinterpret some findings from Section \ref{sec2.7} in a framework more similar to Section \ref{sec3}. To this end, we postulate that the borrowing account $B^b$ satisfies
\[
dB_t^b=f^b_tB^b_t\,dt-L_IB^b_{t-}d\ind{t\geq\tau_I}.
\]
This models the fact that at the bank's default only a fraction $R_I=1-L_I$ of the bank's debt will be repaid to its creditors. Since we are including default in our model, we will also make use of equations \eqref{wtAC} and  \eqref{wtACo} to define the stream of cash flows generated by the contract. In order to value a contract that cannot be replicated, we need to extend slightly the definition of trading strategy of Section \ref{sec3} by including the contract itself. To be more specific, we are interested in buy-and-hold strategies $\phi$ of the form $(1,\psi^l_t,\psi^b_t,\psi^1,\dots,\psi^d,\xi^1,\dots,\xi^d)$ where $1$ indicates one unit of the derivative asset held by the bank. In essence, our valuation arguments in this subsection are based on the idea that the extended market model where the contract is traded should remain arbitrage-free. To be more specific, we will assume that the contract is already held by the bank and we search for the process $\pi$ such that the bank's deflated wealth has the martingale property under a chosen martingale measure. Hence the value $\pi_t$ is here understood as the price at which the bank would be ready to sell the contract at any given date $t$. Therefore, the convention regarding the sign of $\pi$ is consistent with the one previously adopted in Section \ref{sec2}, rather than the convention of Section \ref{sec3} where we examined the replication-based valuation for the contract that could be entered into by the bank at time $t$.

We denote by $W_t(\phi,A^{C,R})$ and $W^p_t(\phi,A^{C,R})$ the wealth and the portfolio's value, respectively, so that on the event $\{t<\tau\}$,
\begin{equation}  \label{buyhold}
W^p_t(\phi,A^{C,R})=\pi_t+\psi^l_tB^l_t+\psi^b_tB^b_t+\sum_{i=1}^d(\psi_t^iB_t^i+\xi_t^iS_t^i) =
\pi_t+\psi^l_tB^l_t+\psi^b_tB^b_t = \pi_t +F_t
\end{equation}
and $W_t(\phi,A^{C,R})= W^p_t(\phi,A^{C,R})-C_t$ where $\pi_t$ is the shorthand for $\pi_t(A,C,R,\tau)$ and where we used the equality $\psi_t^iB_t^i+\xi_t^iS_t^i=0$. Recall that the convention regarding the collateral is that $C_t>0$ means that the bank receives the collateral, while $C_t<0$ that the bank posts it. Hence the minus sign in front of the collateral account is needed since the wealth $W_t(\phi,A^{C,R})$ is what the bank would get at time $t$ if the bank were to liquidate every asset owned. At default time the portfolio's value and the wealth satisfy, on the event $\{\tau\leq T\}$,
\begin{equation}  \label{buyholdtau}
W^p_{\tau}(\phi,A^{C,R})= W_{\tau}(\phi,A^{C,R})=R_\tau+\psi^l_{\tau}B^l_{\tau}+\psi^b_{\tau}B^b_{\tau}.
\end{equation}
The gains process for this strategy is given by, on $[0,T]$,
\begin{align*}
G_t=\ind{t<\tau}\pi_t+\int_0^t\psi^l_u\,dB^l_u+\int_0^t\psi^b_{u-}\,dB^b_u
+\sum_{i=1}^d\left(\int_0^t\xi_u^i\,dS_u^i+\int_0^t\psi^i_u\,dB^i_u\right)+\int_0^t dA_u^{C,R}- \I_{\{t < \tau \}} C_t
\end{align*}
where we define
\[
dA_u^{C,R}=d\widetilde{A}_u+ d\widetilde{C}_u - \ind{u<\tau}\bar{c}_u C_u \, du+d(\I_{\{u \geq \tau \}} R_{\tau})
\]
and we recall that $\widetilde{A}_t=\I_{\{t < \tau \}} A_t + \I_{\{t \geq \tau \}} A_{\tau-}$ and
$\widetilde{C}_t=\I_{\{t < \tau \}} C_t + \I_{\{t \geq \tau \}} C_{\tau-}$. We consider only self-financing strategies, that is, those that satisfy on $[0,T]$
\[
W_t(\phi,A^{C,R})=W_0(\phi,A^{C,R})+ G_t - G_0.
\]
For the deflated wealth $W_t^\eta(\phi,A^{C,R}):=(B_t^\eta)^{-1}W_t(\phi,A^{C,R})$, we get
\[
W_t^\eta(\phi,A^{C,R})=W_0^\eta(\phi,A^{C,R})+ G^\eta_t - G^\eta_0
\]
where
\begin{align*}
dG_u^\eta&=(B_u^\eta)^{-1}\,d\pi_u+(B_u^\eta)^{-1}\psi^l_u\,dB^l_u+(B_u^\eta)^{-1}\psi^b_{u-}\,dB^b_u+\sum_{i=1}^d(B_u^\eta)^{-1}(\xi_u^i\,dS_u^i+\psi^i_u\,dB^i_u)\\
&-(B_u^\eta)^{-1}\ind{u<\tau}\bar{c}_uC_u\,du+(B_u^\eta)^{-1}\,d\widetilde{A}_u+(B_u^\eta)^{-1}\,d(\ind{t\geq \tau}C_{\tau-})+(B_u^\eta)^{-1}\,d(\ind{t\geq \tau}R_{\tau })\\
&-\eta_u(B_u^\eta)^{-1}(\pi_u+\psi_u^lB^l_u+\psi_u^bB^b_u)\,du+ \eta_u(B_u^\eta)^{-1}\ind{u<\tau}C_u\,du .
\end{align*}

To proceed further, we henceforth postulate that the process $G^\eta$ is a martingale with respect to the filtration $\gg$ under a probability measure $\Q^h$, which is a special case of a measure $\Q^{\zeta , \gamma ,\nu}$ introduced in Definition \ref{defmar}.
To be more explicit, a probability measure $\Q^h$ is characterized by the property that the processes $\bar S^i = (B^h)^{-1}S^i,\, i=1,2,\dots ,d $ or, equivalently, the processes
\begin{equation}  \label{stocki}
d\bar S^i_t = (B_t^h)^{-1} ( dS^i_t - h_t S^i_t \, dt )
\end{equation}
are $\Q^h$-local martingales.

\brem Since in this section we deal with an incomplete market model, the uniqueness of
a probability measure $\Q^h$ fails to hold. Therefore, it should be acknowledged that our tentative valuation results presented in what follows hinge in fact on a choice of a particular measure $\Q^h$.
\erem

The postulated martingale property of $G^\eta$ under $\Q^h$ yields, for $t<\widehat {\tau} = \tau \wedge T $,
\[
\EQh ( G_{T\wedge \tau}^\eta-G_{t\wedge \tau}^\eta \,|\,\G_t)
=\EQh ( G_{\widehat {\tau}}^\eta-G_{t}^\eta \,|\,\G_t) =0
\]
and thus we get, on the event $\{t<\widehat {\tau}\}$,
\begin{align*}
(B_t^\eta)^{-1}\pi_t&=\EQh \bigg(\int_t^{\widehat {\tau}}(B_u^\eta)^{-1}\psi^l_u\,dB^l_u
+\int_t^{\widehat {\tau}}(B_u^\eta)^{-1}\psi^b_{u-}\,dB^b_u+\sum_{i=1}^d\int_{(t,\widehat {\tau}]}(B_u^\eta)^{-1}
(\xi_u^i\,dS_u^i+\psi^i_u\,dB^i_u) \,\Big|\,\G_t \bigg) \\
&+\EQh \bigg(\int_{(t,\widehat {\tau}]} (B_u^\eta)^{-1}\,d\widetilde{A}_u-
\int_t^{\widehat {\tau}}(B_u^\eta)^{-1}\bar{c}_uC_u\,du+\ind{\tau<T}(B_\tau^\eta)^{-1}(R_{\tau }+C_{\tau-}) \,\Big|\,\G_t \bigg)\\
&-\EQh \bigg(\int_t^{\widehat {\tau}}\eta_u\psi_u^lB^l_u\,du+\int_t^{\widehat {\tau}}\eta_u\psi_u^bB^b_u\,du
-\int_t^{\widehat {\tau}}\eta_u(B_u^\eta)^{-1}C_u\,du  \,\Big|\,\G_t \bigg)
\end{align*}
where we use the shorthand notation $\pi_t$ for the process $\pi^{\eta,h}_t(A,C,R,\tau)$.
We substitute the expressions for the differentials and use \eqref{closeout} as well as the equality $\xi_u^iS_u^i+\psi_u^iB_u^i=0$ to obtain, on the event $\{t<\widehat {\tau}\}$,
\begin{align*}
(B_t^\eta)^{-1}\pi_t&=\EQh \bigg(\int_t^{\widehat {\tau}}(B_u^\eta)^{-1}(\bar{f}_u-\eta_u)F_u\,du
+\int_{(t,\widehat {\tau}]}(B_u^\eta)^{-1}\,d\widetilde{A}_u
+\int_t^{\widehat {\tau}}(B_u^\eta)^{-1}(\eta_u-\bar{c}_u)C_u\,du \,\Big|\,\G_t \bigg)\\
&+\EQh \Big(\ind{\tau<T}(B_{\tau}^\eta)^{-1}(Q_{\tau } + \I_{\{\tau = \tau_I \}}\big( \lgd_I (Q_{\tau_I}-C_{\tau_I-})^--L_I\psi^b_{\tau_I-}B^b_{\tau_I-} \big) \,\big|\,\G_t \Big)\\
&-\EQh \Big(\ind{\tau<T}(B_{\tau}^\eta)^{-1}\I_{\{\tau = \tau_C \}} \lgd_C (Q_{\tau_C}-C_{\tau_C-})^+ \,\big|\,\G_t \Big)
\end{align*}
where
\[
F_t=\psi_t^lB_t^l+\psi_{t}^bB_t^b=(W_t-\pi_t+C_t)^+-(W_{t}-\pi_{t}+C_{t})^-=F^+_t-F^-_t
\]
and $W_t=W_t(\phi,A^{C,R})$. Equation \eqref{buyhold} gives, on $\{t<\widehat {\tau}\}$,
\begin{align} \label{piminus}
&(B_t^\eta)^{-1}\pi_t=\EQh \bigg(\int_t^{\widehat {\tau}}(B_u^\eta)^{-1}(\bar{f}_u-\eta_u)\big((W_u-\pi_u+C_u)^+-(W_{u}-\pi_{u}+C_{u})^-\big)\,du \,\Big|\,\G_t \bigg) \nonumber \\
&+\EQh \bigg(\int_{(t,\widehat {\tau}]} (B_u^\eta)^{-1}\,d\widetilde{A}_u+\int_t^{\widehat {\tau}}(B_u^\eta)^{-1}(\eta_u-\bar{c}_u)C_u\,du  \,\Big|\,\G_t \bigg)\\
&+\EQh \Big(\ind{\tau<T}(B_{\tau}^\eta)^{-1}\big(Q_{\tau } + \I_{\{\tau = \tau_I \}} \lgd_I (Q_{\tau_I}-C_{\tau_I-})^- -  \I_{\{\tau = \tau_C \}} \lgd_C (Q_{\tau_C}-C_{\tau_C-})^+ \big) \,\big|\,\G_t \Big) \nonumber \\
&+\EQh \Big( \ind{\tau<T}\ind{\tau=\tau_I}\lgd_I(B_{\tau-}^\eta)^{-1}(W_{\tau-}-\pi_{\tau-}+C_{\tau-})^- \,\big|\,\G_t \Big). \nonumber
\end{align}
A priori, the right-hand side in the above tentative pricing formula depends on the wealth process $W$ of the chosen strategy $\phi$. However, if we set
$$
\bar{f}_t=\eta_t\ind{F_t>0}+ (s^f_t + \eta_t) \ind{F_t\leq 0}
$$
where $s^f_t$ is the {\it funding spread} of the bank with respect to the prevailing deposit rate $\eta_t$,  then it is clear that the first and last terms in the right-hand side of equation \eqref{piminus} may cancel out if the funding spread is `fair', that is, if the bank's default and recovery are priced correctly by the market.

To examine the last statement in more detail, we denote $Y_t= (B_t^\eta)^{-1}(W_{t}-\pi_{t}+C_{t})^-$ and we examine the
{\it net funding/default benefit} $\DVAFn_t-\FCA^f_t$, which in the present setup is given by the following expression
\[
J_t:= \EQh \big(\ind{\tau<T} \ind{\tau=\tau_I}\lgd_I Y_{\tau-}\,|\,\G_t \big)- \EQh \bigg( \int_t^{\widehat \tau } s^f_u Y_u \, du\,\,\Big|\,\G_t \bigg).
\]
Our goal is to provide explicit conditions under which the bank's net funding benefit vanishes. To this end, we postulate that  $\lgd_I$ is a constant and default times $\tau_I$ and $\tau_C$ are conditionally independent under $\Q^h$ with respect to the reference filtration $\ff$ (see, for instance,  Example 9.1.5 in \cite{BR02}) with stochastic intensities $\lambda^I$ and $\lambda^C$, respectively. Then the intensity of $\tau = \tau_I \wedge \tau_C$ satisfies $\lambda = \lambda^I+\lambda^C$ and we can use the standard intensity-based approach to complete computations of $J_t$ (see, for instance, Propositions 5.1.1 and 5.1.2 in \cite{BR02} and
Lemma 3.8.1 in \cite{BJR09}). In particular, we obtain
\begin{align*}
J_t&=\lgd_I \, \EQh \big(\ind{\tau<T} \ind{\tau=\tau_I} Y_{\tau_I-}\,|\,\G_t \big)- \EQh \bigg( \int_t^{\tau \wedge T} s^f_u Y_u \, du\,\,\Big|\,\G_t \bigg) \\
&= \ind{t<\tau } \, \EQh \bigg( \int_t^{T} e^{\Lambda_t-\Lambda_u }\lgd_I\lambda^I_u Y_u \, du \,\Big|\,\F_t \bigg) -\ind{t<\tau} \EQh \bigg( \int_t^{T} e^{\Lambda_t-\Lambda_u} s^f_u Y_u \, du  \,\Big|\,\F_t \bigg)
\end{align*}
where the hazard process $\Lambda $ of $\tau$ satisfies $\Lambda_t = \int_0^t \lambda_u \, du $. It is thus clear that if $s^f_t = \lgd_I \lambda^I_t$ for all $t\in [0,T]$, then $J_t=0$ for all $t \in [0,T]$ and thus the equality $\FCA^f_t = \DVAFn_t$ is satisfied for all $t \in [0,T]$. This shows that, in some circumstances, the funding cost may  completely offset the benefit at default (see Section \ref{sec2.7} for similar considerations).

We conclude this paper by stating the proposition, which can be seen as an extension of Lemma \ref{lemm2.1} and is also
related to Proposition \ref{profin}. Recall that we work under the postulate that for any self-financing strategy the process $G^\eta$ is a martingale under $\Q^h$. We denote
$$
\pi^{\eta ,h}_t (A) :=B_t^\eta \, \EQh \bigg(\int_{(t,T]}(B_u^\eta)^{-1}\,dA_u  \,\Big|\,\G_t \bigg)
$$
and
$$
Q^{\eta ,h}_t :=B_t^\eta \, \EQh \bigg(\int_{[t,T]}(B_u^\eta)^{-1}\,dA_u  \,\Big|\,\G_t \bigg)=\Delta A_{\tau}+\pi^{\eta ,h}_{\tau}(A).
$$
The following result shows that the double counting of benefits arising due to a possibility of the bank's default
does not appear if the bank's trading arrangement are modeled adequately, although the net funding/default benefit
does not necessarily vanish.

\bp \label{pro4.1}
Assume that $\lgd_I$ is a constant and default times $\tau_I$ and $\tau_C$ are conditionally independent under $\Q^h$ with respect to the reference filtration $\ff$. Let the intensity of the bank's default time $\tau_I$ under $\Q^h$ with respect to the filtration $\ff$ equal $\lambda^I $ and let for all $t\in [0,T]$
$$
\bar{f}_t=\eta_t\ind{F_t>0}+ (\lgd_I \lambda^I_t + \eta_t) \ind{F_t\leq 0}
$$
so that the bank's funding spread equals $s^f_t= \lgd_I \lambda^I_t$ for all $t\in [0,T]$.
Then {\rm $\FCA^f_t=\DVAFn_t$} on the event $\{t< \tau \}$ so that the net funding/default benefit vanishes and the ex-dividend selling price $\pi^{\eta,h}_t(A,C,R,\tau)$ satisfies
\begin{align*}
&\pi^{\eta,h}_t(A,C,R,\tau)= \pi^{\eta ,h}_t (A)+B_t^\eta \, \EQh \bigg(\int_t^{\tau \wedge T}(B_u^\eta)^{-1}(\eta_u-\bar{c}_u)C_u\,du  \,\Big|\,\G_t \bigg)\\
&+B_t^\eta \,\EQh \Big(\ind{\tau<T}(B_{\tau}^\eta)^{-1}\big( \I_{\{\tau = \tau_I \}} \lgd_I (Q^{\eta,h}_{\tau_I}-C_{\tau_I-})^- -  \I_{\{\tau = \tau_C \}} \lgd_C (Q^{\eta,h}_{\tau_C}-C_{\tau_C-})^+ \big) \,\big|\,\G_t \Big)
\end{align*}
where $Q^{\eta ,h}_{\tau}=\Delta A_{\tau}+\pi^{\eta ,h}_{\tau}(A)$. Hence the price $\pi^{\eta,h}_t (A,C,\recR,\tau )$ admits the following representation
{\rm 
\begin{align*}
\pi^{\eta,h}_t (A,C,\recR,\tau )= \pi^{\eta ,h}_t (A)+\LVA_t+\DVA_t-\CVA_t.
\end{align*} }
\ep

\proof
Using \eqref{piminus} and noting that, under the present assumptions, we have that $J_t=0$ for all $t\in [0,T]$, we obtain
\begin{align*}
&\pi^{\eta,h}_t(A,C,R,\tau)= B_t^\eta \, \EQh \bigg(\int_{(t,\tau \wedge T ]}(B_u^\eta)^{-1}\,d\widetilde{A}_u+\int_t^{\tau \wedge T}(B_u^\eta)^{-1}(\eta_u-\bar{c}_u)C_u\,du  \,\Big|\,\G_t \bigg)\\
&+\EQh \Big(\ind{\tau<T}(B_{\tau}^\eta)^{-1}\big(Q_{\tau } + \I_{\{\tau = \tau_I \}} \lgd_I (Q_{\tau_I}-C_{\tau_I-})^- -  \I_{\{\tau = \tau_C \}} \lgd_C (Q_{\tau_C}-C_{\tau_C-})^+ \big) \,\big|\,\G_t \Big).
\end{align*}
If we postulate that $Q = Q^{\eta ,h}$, then to obtain the desired representation for $\pi^{\eta,h}_t(A,C,R,\tau)$, it suffices to proceed as in the proof of Proposition \ref{pro1.2}.
\endproof

Observe that, under the assumptions of Proposition \ref{pro4.1}, the ex-dividend selling price $\pi^{\eta,h}(A,C,R,\tau)$ is independent of the wealth process $W$ of the chosen strategy $\phi$. Note also that $\DVA_t$ and $\CVA_t$ are always nonnegative. However, the sign of $\LVA_t$ depends on the relationship between $\eta$ and $\bar{c}$ as well as the sign of $C$.



\end{document}